\documentclass[10pt, conference, compsocconf]{IEEEtran}
\usepackage{mathptmx} % This is Times font

\usepackage{fancyhdr}
\usepackage[normalem]{ulem}
\usepackage[hyphens]{url}
\usepackage[sort,nocompress]{cite}
\usepackage[final]{microtype}
\usepackage[keeplastbox]{flushend}
\usepackage[bookmarks=true,breaklinks=true,letterpaper=true,colorlinks,citecolor=blue,linkcolor=blue,urlcolor=blue]{hyperref}

\usepackage{comment}
\usepackage{tikz}
\usepackage{graphicx}
\usepackage{color}
\usepackage{tabularx}
\usepackage{bchart}
\newcommand{\hyeran}[1]{\textcolor{magenta}{#1}}

\usepackage{subfig}
\usepackage{multirow}
\usepackage{algorithm} 
\usepackage{algpseudocode} 
\usepackage{soul}

\begin{document}

%%%%%%%%%%%---SETME-----%%%%%%%%%%%%%
\title{Revealing Secrets From Pre-trained Models} 
%%%%%%%%%%%%%%%%%%%%%%%%%%%%%%%%%%%%
\author{
\IEEEauthorblockN{Mujahid Al Rafi, Yuan Feng, Hyeran Jeon}
\IEEEauthorblockA{University of California, Merced
\\\{mrafi, yfeng44, hjeon7\}@ucmerced.edu}
}

\maketitle

%\pagestyle{plain}

%%%%%% -- PAPER CONTENT STARTS-- %%%%%%%%

\begin{abstract}
With the growing burden of training deep learning models with large data sets, transfer-learning has been widely adopted in many emerging deep learning algorithms. Transformer models such as BERT are the main player in natural language processing and use transfer-learning as a de facto standard training method. A few big data companies release pre-trained models that are trained with a few popular datasets with which end users and researchers fine-tune the model with their own datasets. Transfer-learning significantly reduces the time and effort of training models. However, it comes at the cost of security concerns. In this paper, we show a new observation that pre-trained models and fine-tuned models have significantly high similarities in weight values. Also, we demonstrate that there exist vendor-specific computing patterns even for the same models. With these new findings, we propose a new model extraction attack that reveals the model architecture and the pre-trained model used by the black-box victim model with vendor-specific computing patterns and then estimates the entire model weights based on the weight value similarities between the fine-tuned model and pre-trained model. We also show that the weight similarity can be leveraged for increasing the model extraction feasibility through a novel weight extraction pruning.
\end{abstract}

\section{Introduction}
Machine learning has been rapidly adopted to virtually almost all the computing fields from image recognition to natural language processing. The accuracy and performance of machine learning models determine the solution providers' market revenue. To acquire startups that provide outstanding machine learning solutions, big tech companies spend billions of dollars recently. Therefore, the machine learning model implementation details are one of the top priority secrets that the solution providers need to protect. However, recent a few studies demonstrated that it is feasible to steal model topology, hyperparameters, and training dataset~\cite{deepsniffer_asplos20, hoda_ccs18, mengjia_security20}. These attack models, namely \textit{model extraction attack}, \textit{hyperparmeter stealing attack}, and \textit{membership inference attack} leverage side channels through performance counters, cache access timings, and bus probing. Most of the existing attack models targeted models used for image processing such as convolutional neural network (CNN). %because CNNs use fairly regular network structure and most of the open-source frameworks commonly use handful of library functions such as OpenBLAS or cuDNN. 
Though revealing model architectures is possible, it is known to be very challenging to reveal actual weight values due to the abundant size. Some recent studies revealed one- to eight-bit weight values from ResNet model with an extended RowHammer attack method~\cite{deepsteal}. To recover 80 $\sim$ 90\% accurate weight values, they had to run at least 4000 hammer-leak rounds.

\begin{figure}
\centering
  \includegraphics[width=0.45\textwidth]{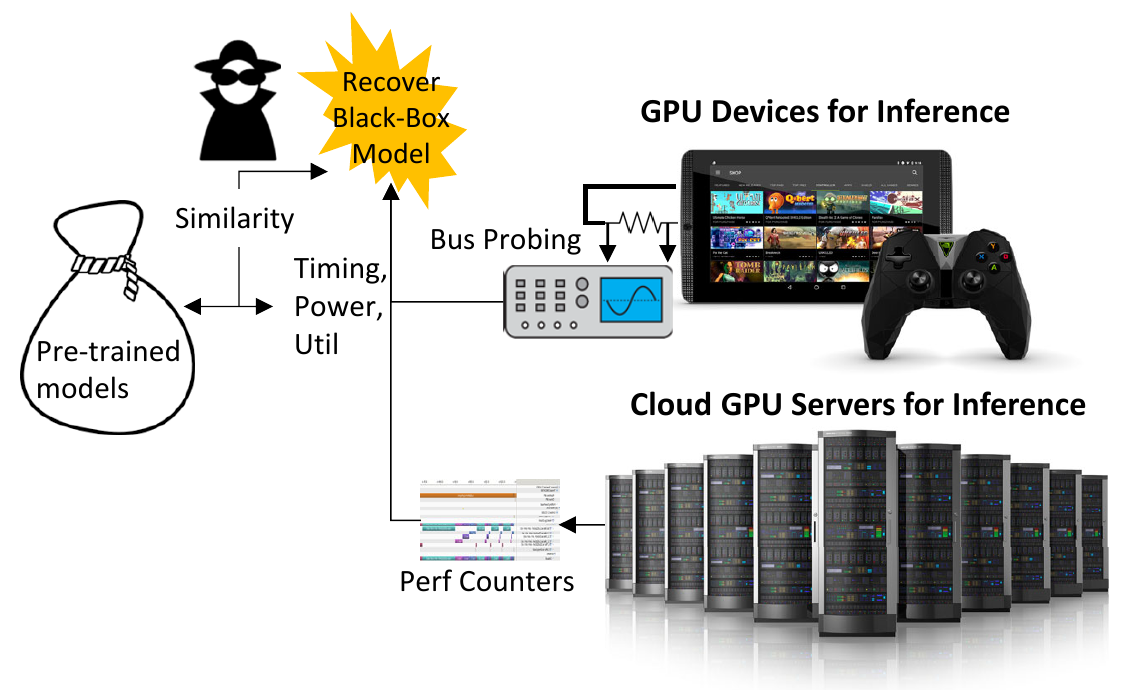}
\caption{Proposed Attack Model} \label{fig.overview}\vspace{-15pt}
\end{figure}

In this paper, we show an easier way to reveal almost accurate weight values as well as model structure by exploiting a recent model training trend. To save model training time and also develop more accurate models with larger dataset, transfer-learning has been widely adopted. Transformer model and its variants such as BERT~\cite{devlin2018bert} and GPT~\cite{gpt-2, gpt-3} follow transfer-learning. A few big data companies release general-purpose models that are pre-trained with large un-labeled datasets and individual developers and other companies develop their own model by fine-tuning the pre-trained model with their own datasets. %We can find tens of thousands of pre-trained models released by Google, Meta, Huggingface, and NVIDIA and many more. 
Due to the obvious cost and performance benefit of transfer-learning, it is very rare to develop these models from scratch. For example, the pre-trained BERT-base and GPT-2 models released on the Huggingface model repository have been downloaded more than 10 million times and 20 million times, respectively~\cite{bert-base-pretrain, gpt-pretrain}. %, It gathers over 30000 pretrained and finetuned models, and among them, BERT-Base\footnote{\url{https://huggingface.co/bert-base-uncased}} model is downloaded more than 10 million times and the most downloaded model DistilGPT2\footnote{\url{https://huggingface.co/distilgpt2}} has been downloaded for over 20 million times. Due to the high pretraining cost and the accessibility of the rich public resource, the pretrain-finetune has became the new ML paradigm.
However, we observed that the pre-trained models are double-edged sword that not only provides model development efficiency but also exposes model secrets unintentionally. From our observation, many fine-tuned models show very high similarities in weight values with the baseline pre-trained models. Once an attacker grabs the information about the baseline pre-trained model with various adversarial attack methods~\cite{deepsniffer_asplos20, mengjia_security20, hoda_ccs18}, the black-box fine-tuned model's weight values can be easily estimated without thousands to millions of memory probing. 

Regarding the adversarial attacks, unlike existing studies assumed, we observed that system statistics do not show unique pattern for individual models. Even for the models that use the same architecture and are trained with the same dataset for the same task (e.g., BERT-large for question-answering task trained with SQuAD dataset), the system statistics show notable variances if the models are released by different vendors and use different frameworks. %We observed that the diversity is rooted from the disparity of vendor (that designed the pre-trained model) and framework (e.g., TensorFlow, PyTorch, etc). 
We observed that individual vendors tend to use unique sets of GPU kernels mainly due to different framework usage and optimization preferences, which lead to diverse system statistics. Therefore, we found that the existing model extraction attack methods show good extraction accuracy only when the victim model uses exactly the same GPU kernels that the adversary developed the attack model with. On the other hand, we found that such vendor/framework signatures are a good source of information about the pre-trained model that the black-box model is developed with. As a pre-trained model and its fine-tuned model have very similar weight values, once the baseline pre-trained model is identified, almost similar weights can be revealed in a whole model level. 

In this paper, we present a new model extraction attack that leverages the vendor/framework signatures and weight value similarity to extract model architecture as well as weight values. We also propose weight extraction pruning that significantly reduces the memory probing and hammering scopes in finding the exact weight values by leveraging the uncovered pre-trained model's weights and the typical value distances between pre-trained models and their fine-tuned models. In summary, we show an easier and more accurate model extraction attack for transfer-learning based models. We show examples and evaluations with mostly BERT models but the ideas can be extended to any models that use transfer-learning, as we show a CNN example in Section~\ref{sec:experiment}. Figure~\ref{fig.overview} illustrates the proposed attack model where an adversary uses not only system statistics but also patterns found from pre-trained models for recovering a black-box model. 
%From the signatures that can be monitored via performance counters and execution times, we were able to reveal the baseline pre-trained models from a black-box fine-tuned models. 

Our contributions are like below:

\begin{enumerate}
    \item We show novel observations that can be leveraged for developing a more accurate model extraction attack for transfer-learning based models: 1) high value similarity between a pre-trained model and its fine-tuned model and 2) vendor/framework signature in architecture statistics
    \item We develop a new model extraction attack that recovers not only the victim model architecture but also the entire (almost similar) weight values by exploiting the aforementioned two observations. With our proposed weight extraction pruning, the actual weight values can be more easily recovered. 
    \item To our best knowledge, this is the first study that extracts architecture of Transformer models. Our observation shows that the architectural uniqueness of Transformer models eases the extraction and hence an adversary does not need to check multiple side-channels; only the GPU-side function execution time monitoring is sufficient. 
\end{enumerate}
\vspace{-5pt}
\begin{figure}[t]
\centering
  \includegraphics[width=0.45\textwidth]{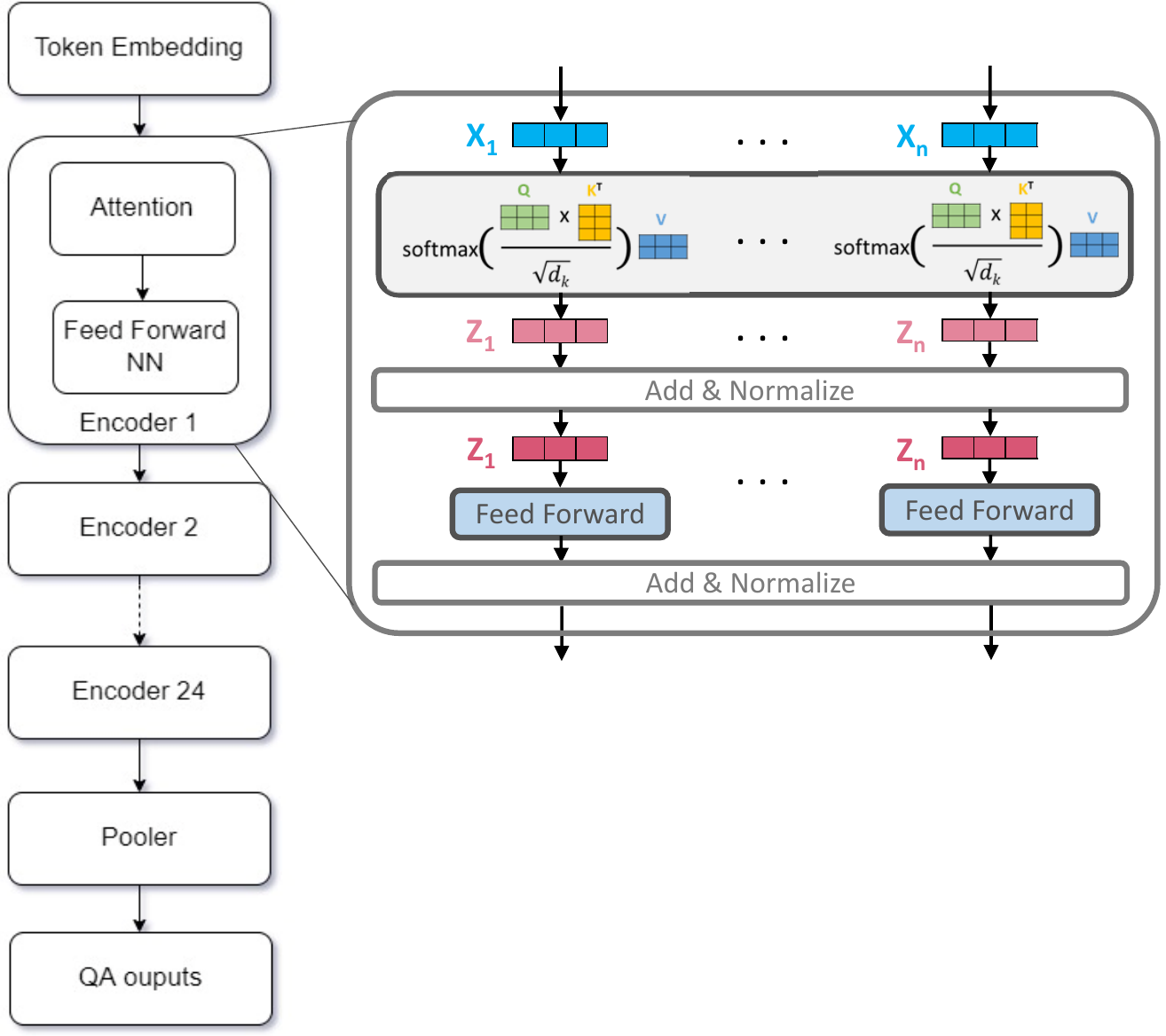}
\caption{BERT-large Architecture} \label{fig.bertarch}\vspace{-15pt}
\end{figure}
%The rest of paper is organized like below:

\section{Transformer-based Models}
Though transfer-learning can be technically used by any machine learning algorithms, one of the most popular models that use transfer-learning as default training method is Transformer model. Transformer models have proved their effectiveness in various domains including natural language processing (NLP) and computer vision~\cite{attention_17}. Especially in NLP, Transformer is efficient to be adapted to several diverse tasks such as question answering, sentiment analysis, named entity recognition, etc with a task-specific last layer attached to the baseline pre-trained model architecture. In spite of large model size, Transformer models gain popularity due to its parallel architecture. Unlike earlier NLP solutions that relied on recurrent and convolution computations, Transformer models focus on attention computations. By extracting relations across input tokens, Transformer locates the most influential parts of the input that determine the output prediction. A Transformer model runs multiple rounds of attention computations where each attention layer checks relations across all tokens of the given input in parallel. %Because many Transformer-based models and their variants are proposed and studied by prior works, the base-line transformer proposed in \cite{attention_17} is used for introduction purpose.

Transformer models can consist of one (or both) of two key modules: \textit{encoder} and \textit{decoder}. 
Each \textit{encoder} is further comprised of two layers: self-attention and feed-forward, as illustrated in Figure~\ref{fig.bertarch}. The unique algorithmic strength of Transformer models is sourced from the self-attention layer. %The key mechanism of transformer model is the self-attention module. 
In a self-attention layer, properly embedded input tokens are multiplied with three weight matrixes, namely Key ($W_k$), Query ($W_q$), and Value ($W_v$) and generate K, Q, and V vectors. These three vectors for each token are used for understanding the importance of the token in the input context. %the hidden states that are embedded from the input sequences are fed into three different linear production modules, namely Key, Query, and Value. Each of the three modules then generates an intermediate variable vector, $K$, $Q$ and $V$. 
Then \textit{attention scores} are calculated via a multiplication of $Q$ vector and $K$ vector followed by a normalization ($\alpha$) and a softmax operation. Then, the attention score is applied to $V$ vector to calculate a weighted Value, $V'$. $V'$s are accumulated to a $Z$ vector and passed to a feed-forward layer, and then to the next encoder. After going through cascaded encoder layer computations, a task-specific layer computes a classification result to generate a prediction output. %Final results $Z$ are calculate by multiplication of the \textit{attention scores} and the $V$ vector. 
The overall computation is depicted in Equation \ref{eq:1}. %It should be noted that to capture the features in different domain more efficiently, hidden states are further divided into several sub-matrix and process individually, called \textit{attention head}, though, the overall computation keep the same.
\vspace{-5pt}
\begin{equation} \label{eq:1}
Z = Softmax( \frac{Q(i)\cdot K(i)^{T} }{\alpha}) \cdot V(i)
\end{equation}

Decoders share a similar architecture with encoders, except for one key difference. %, where there are two main differences.
While an encoder processes the whole input hidden states at once, a decoder only processes one hidden state in a single forward computation. This leads to a slight difference to self-attention layer where a decoder uses masked self-attention such that future tokens are excluded from the computation. %First, while encoder processes the whole input hidden states at once, the decoder only processes one hidden state in a single forward computation. %Second, $K$ vectors and $V$ vectors are obtained from the encoding phase, while $Q$ vectors are generated during the decoding phase each forward run. 

While Transformer models are introduced to solve problems in machine translation domain, several Transformer variant models are proposed to the other domains with keeping similar model architecture. The variants can be summarized into two groups: 1) discriminative models, such as BERT\cite{devlin2018bert}, RoBERTa\cite{liu2019roberta}, ALBERT\cite{lan2019albert}, which only contain the encoders and 2) generative models, such as GPT-2\cite{gpt-2} and GPT-3\cite{gpt-3} which contain either decoders only or both encoders and decoders. %encoders are used for analyzing context and relative attentions among the tokens of the given input and finding one answer for one of various tasks through a classifier, while decoders are used for generating next possible token from the context of the provided token inputs. The two representative transformer models, BERT and GPT, are optimized for different types of tasks because BERT uses only encoders and GPT uses decoders. BERT is used for finding an answer from the given input context and question, such as question answering and sentence sentiment detection, which is well supported by encoder. On the other hand, GPT fits well for text generation tasks such as next word prediction, which can be implemented with decoders. 
%Among those different pre-trained model with different pre-define model setting are release for public use. 
Though any number of encoders and decoders can be used for building a model, there are several well-evaluated reference models such as BERT-base, BERT-large, GPT-2 medium and large, and GPT-3. Out of them, we use BERT and its variant models as the main attack targets in this paper. The main differences between BERT-base and BERT-large models are the number and size of encoders where BERT-base uses 12 encoders with 12 attention heads each while BERT-large includes 24 encoders with 16 attention heads each. The internal layers of each encoder are the same, as described above. The BERT-large architecture is illustrated in Figure~\ref{fig.bertarch}.

% Each encoder block has a self-attention layer and feed-forward neural network. In self-attention layer, each token embedding is multiplied by three matrices to get Query (Q), Key (K) and, Value (V) matrices. These three matrices are processed to get the self-attention output. This output is then feed to a feed-forward neural network to produce input for the next encoder block. % commented by Yuan

\begin{figure*}[t]
\centering
  \includegraphics[width=1\textwidth]{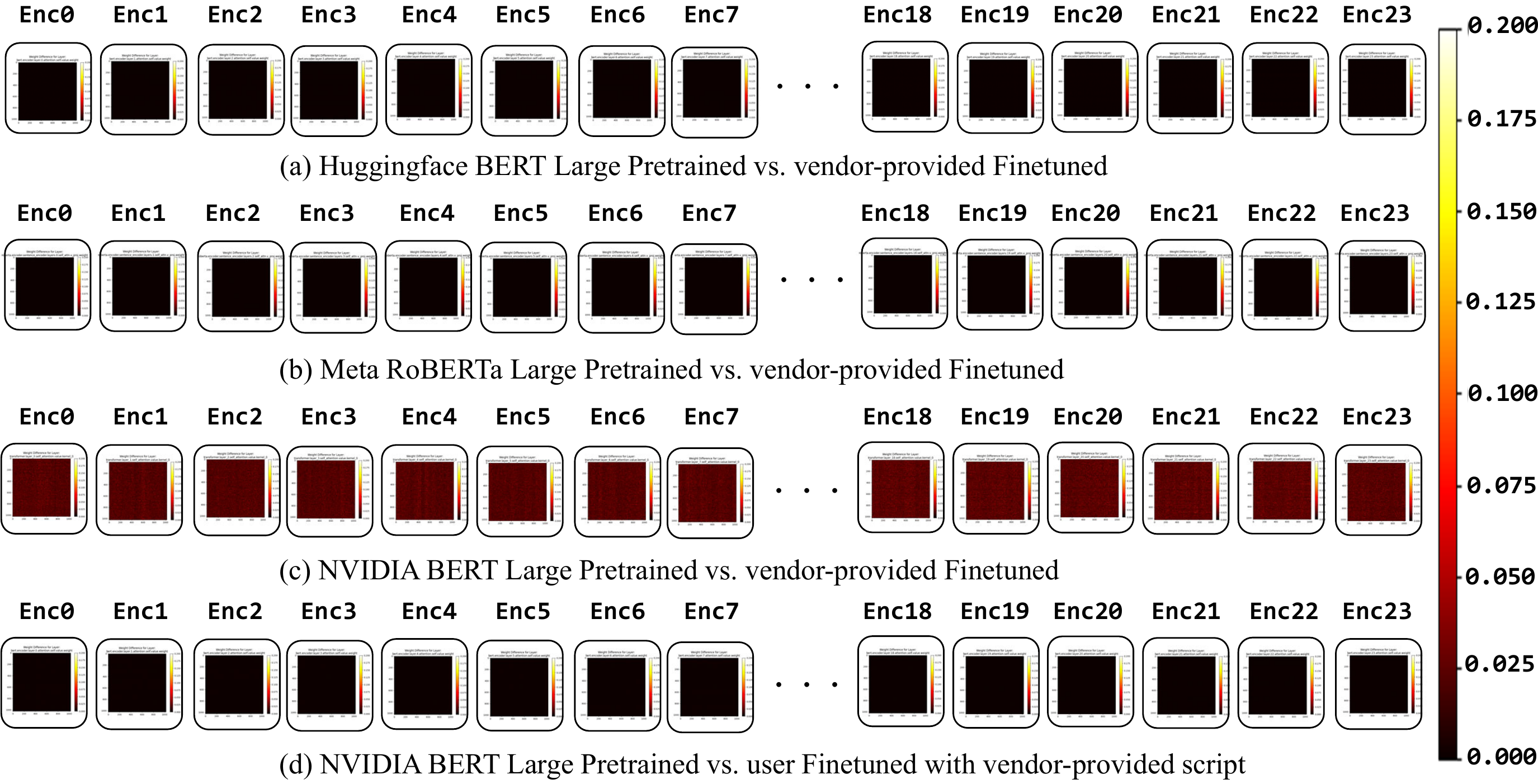}
\caption{Per-Encoder Weight ($W_{v}$) Value Similarities between Pre-trained model vs. Fine-tuned model: weight difference color map on right (black : higher similarity, white : lower similarity) } \label{fig.similarity}
\end{figure*}

\begin{figure*}[h]
\centering
  \includegraphics[width=1\textwidth]{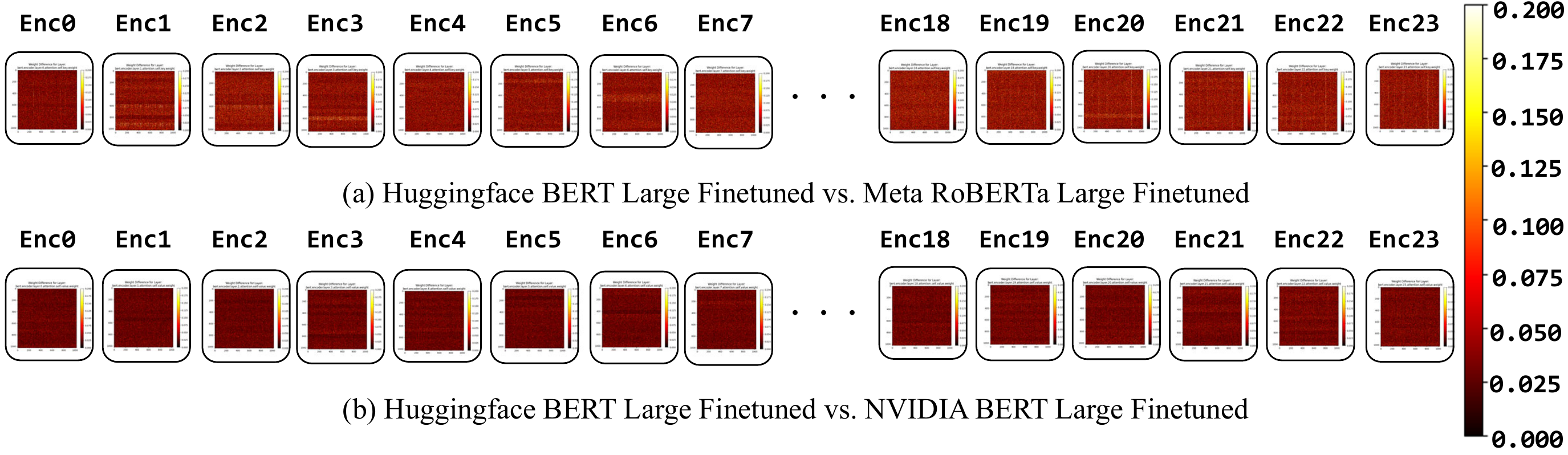}
\caption{Per-Encoder Weight ($W_{v}$) Value Similarities between Different Vendor Models } \label{fig.diff}\vspace{-10pt}
\end{figure*}

\section{Motivations}\label{sec.motivation}\vspace{-5pt}
\subsection{Weight Value Similarity}

To understand the relation between pre-trained models and their fine-tuned models, we compared weight values of four BERT implementations, as can be seen in Figure~\ref{fig.similarity}. We tested the pre-trained and fine-tuned BERT (RoBERTa) Large models released by three vendors, Huggingface~\cite{hf_bert_pt}, Meta~\cite{mt_roberta_pt}, and NVIDIA~\cite{nv_bert_pt, nv_bert_tf}. We compared the absolute value of individual weights of the Value weight matrix ($W_{v}$) used by the self-attention computation in each encoder layer. Each square in the Figure~\ref{fig.similarity} is a value similarity heat map for 1024$\times$1024 $W_{v}$ per encoder. As illustrated on the right-hand side color map, black-colored entries mean the weight value in the same location of each weight matrix of the pre-trained model and the fine-tuned model have high similarity (almost no difference), while the white-colored entries indicate that the value distance between the pre-trained and fine-tuned weights are almost 0.2. Surprisingly, the models of Huggingface and Meta show very minimal value disparity (< 0.002) for all encoder layers. The NVIDIA-released fine-tuned model shows higher differences but it is also not very significant, where the averaged difference is around 0.025. When we fine-tuned the released pre-trained model with the vendor provided fine-tuning script (without changing anything), the weight similarity became higher, like Huggingface and Meta, as can be seen in Figure~\ref{fig.similarity}(d). It seems that NVIDIA might have applied some optimizations over the fine-tuning script for the released fine-tuned models, but again the weight difference was insignificant. Though we show only $W_{v}$ value comparisons results due to the limited space, weights for Query and Key also had similar patterns. 
\begin{figure}[h]
\centering
  \includegraphics[width=0.5\textwidth]{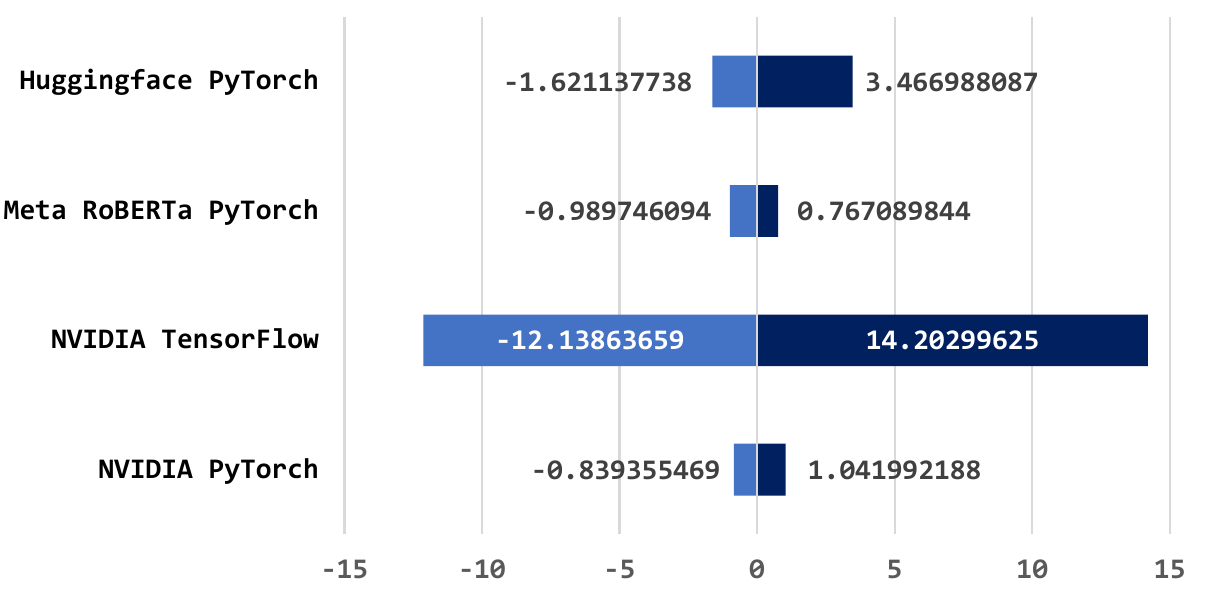}
\caption{Weight Value Range of BERT-large Models} \label{fig.weight_range}\vspace{-15pt}
\end{figure}

\begin{figure}[t]
\centering
\includegraphics[width=0.5\textwidth]{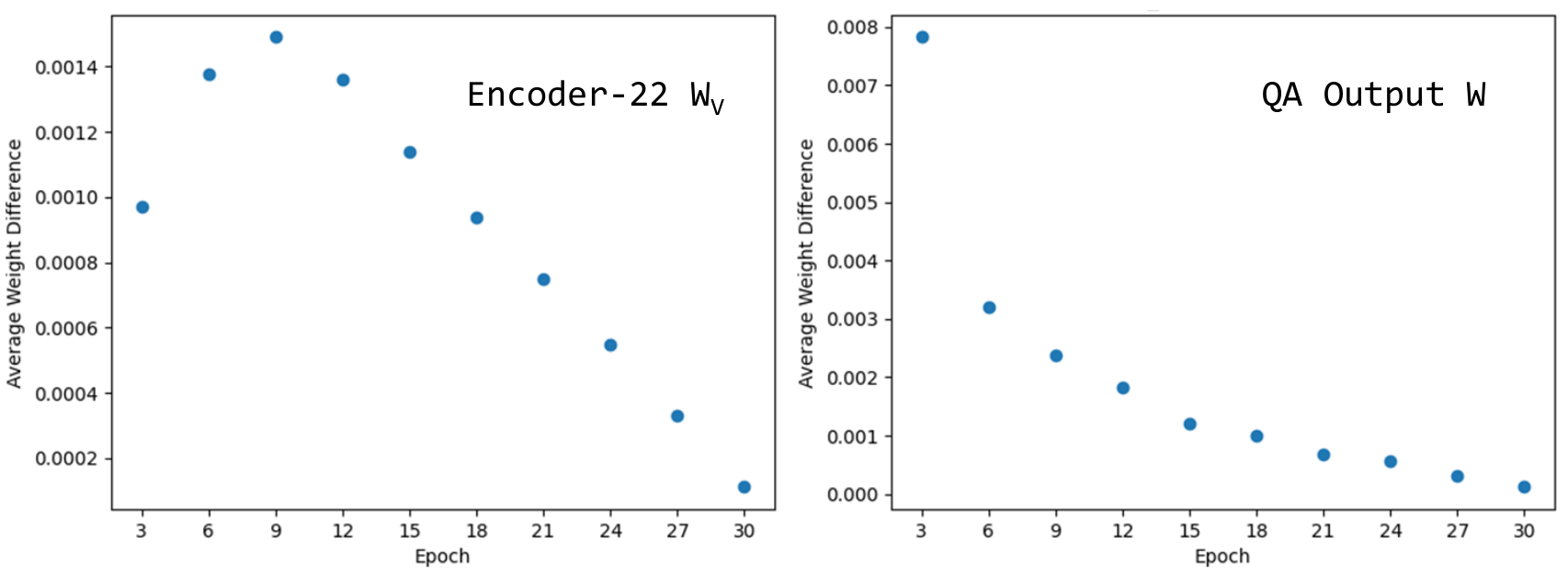}
\caption{Avg. Weight Changes during 30-Epoch Fine-tuning%: examples of W$_v$ of encoder 22 and QA output layer of a BERT-large NVIDIA PyTorch model
} \label{fig.long_finetuning}\vspace{-10pt}
\end{figure}

One might think that the value similarity is sourced from small weight values. However, the weight value ranges of the tested BERT-large models are at least 1.74 up to over 26.3, as can be seen in Figure~\ref{fig.weight_range}. To investigate if such high value similarity is found only between a pre-trained model and its own fine-tuned model, we did another comparison between different vendors-released models, as can be seen in Figure~\ref{fig.diff}. With the same color scale, we observed notably higher differences in this case, where the averaged value gap is around 0.075 $\sim$ 0.1. %From these experiments, we can get to a reasonable conclusion that a pre-trained model and its fine-tuned model are likely to have high weight-value similarity. 
We suspect that fine-tuning hyper-parameter values play some role for this similarity. According to our observations, the fine-tuning scripts provided by all the examined vendors use up to three epochs with small learning rates, which is around 3e-05. Therefore, fine-tuned models might have very subtle differences from the pre-trained models. To understand if the similarity is caused purely by the fine-tuning parameters, we also measured the weight value changes while running longer epochs. Figure~\ref{fig.long_finetuning} shows the average weight value changes of $W_v$ of encoder 22 and task-specific output layer (question-answer task in this example) of an NVIDIA BERT-large PyTorch model collected while fine-tuning a vendor-released pre-trained model for 30 epochs. The weights of $W_v$ have average of 0.001 difference at epoch 3 compared to the pre-trained model, which is also shown in Figure~\ref{fig.similarity}(d). Until epoch 9, weight difference increases up to 0.0015 and then drops linearly to below 0.0002 at epoch 30. In the meantime, the output layer's weights are saturated exponentially as plotted in the second graph in Figure~\ref{fig.long_finetuning}, which means that the fine-tuning converges well. These results show that fine-tuning may achieve better accuracy with more epochs but the weight values do not significantly change even with longer epochs. The reasons can be twofold: 1) pre-trained models are already highly trained with large dataset and 2) fine-tuning is more like an adaptation process for a general-purpose pre-trained model to be used for a specific task. Therefore, the most important step of fine-tuning is not changing the weights in vast majority of the model architecture but attaching a task-specific last layer to a pre-trained model and tuning the weights slightly to fit well for the fine-tuned task. 

In the perspective of performance and prediction accuracy, weight value similarity between pre-trained model and its fine-tuned model may not raise any concerns. However, in security aspect, this fact can be leveraged to extract weight values from a black-box fine-tuned model. Once the model architecture and the vendor is recovered, weight values can be estimated to be within < 0.002 value range from the vendor's publicly accessible pre-trained model. Note that anybody can access pre-trained models as pre-trained models are considered as public resources for transfer-learning models. 

\subsection{Vendor \& Framework Signatures}

\begin{figure}[t]
\centering
\includegraphics[width=0.4\textwidth]{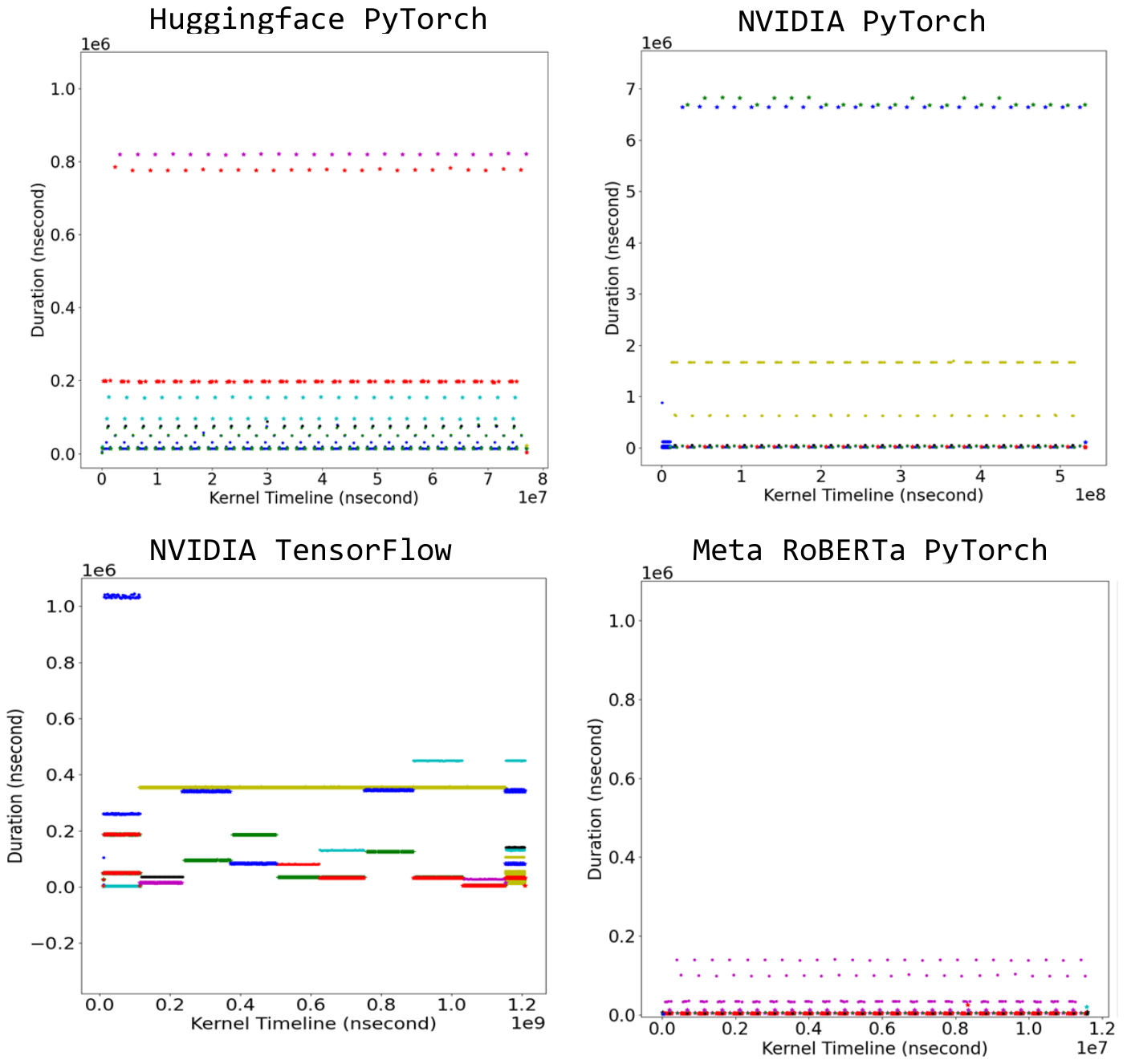}
\caption{Diversity in Time-series Kernel Execution Times for BERT-large Models Released by Different Vendors} \label{fig.vendorvari}\vspace{-10pt}
\end{figure}

\begin{figure}
\centering
\includegraphics[width=0.38\textwidth]{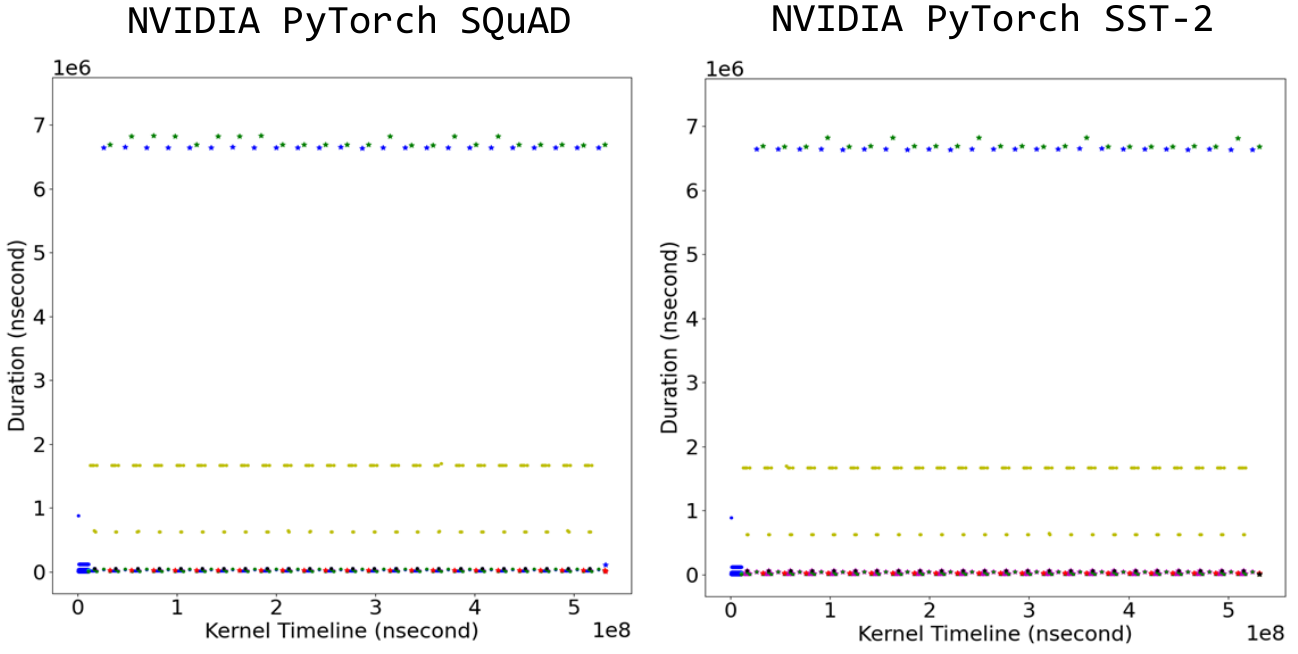}
\caption{Consistency in Time-series Kernel Execution Times for BERT-large Models Trained for Different Tasks Released by One Vendor} \label{fig.vendorcons}\vspace{-15pt}
\end{figure}
To extract victim model architecture, we investigated the common characteristics of various Transformer models such as BERT and RoBERTa. Interestingly, we observed quite diverse statistics from the models released by different vendors even when the models use the same architecture. Figure~\ref{fig.vendorvari} shows time-series execution times of kernels launched by individual models on the same GPU. Each dot indicates the execution time of a kernel and the same-colored dots are the multiple invocations of the same kernel. Each graph plots the statistics of a BERT-large model that is fine-tuned and released by the specified vendor for question-answering task and uses the stated framework. Meta's is RoBERTa-large that basically uses the same model architecture with BERT-large while using different datasets for pre-training. All models are tested with the same input. As can be seen, there are not common patterns found across them. However, for each vendor's models that use the same framework, the statistics show high consistency even when they are fine-tuned for different tasks. Figure~\ref{fig.vendorcons} shows the results of the same test for NVIDIA PyTorch BERT-large models that are fined-tuned for (left) question answering and (right) sentiment analysis. Though there are some diversity in GPU kernel usage, the vast majority of kernels used by both models are identical and the statistics show almost indistinguishable patterns. We observed such vendor-unique statistics for all tested models. We call this as \textit{Vendor Signature}.

\begin{figure}[t]
\centering
  \includegraphics[width=0.45\textwidth]{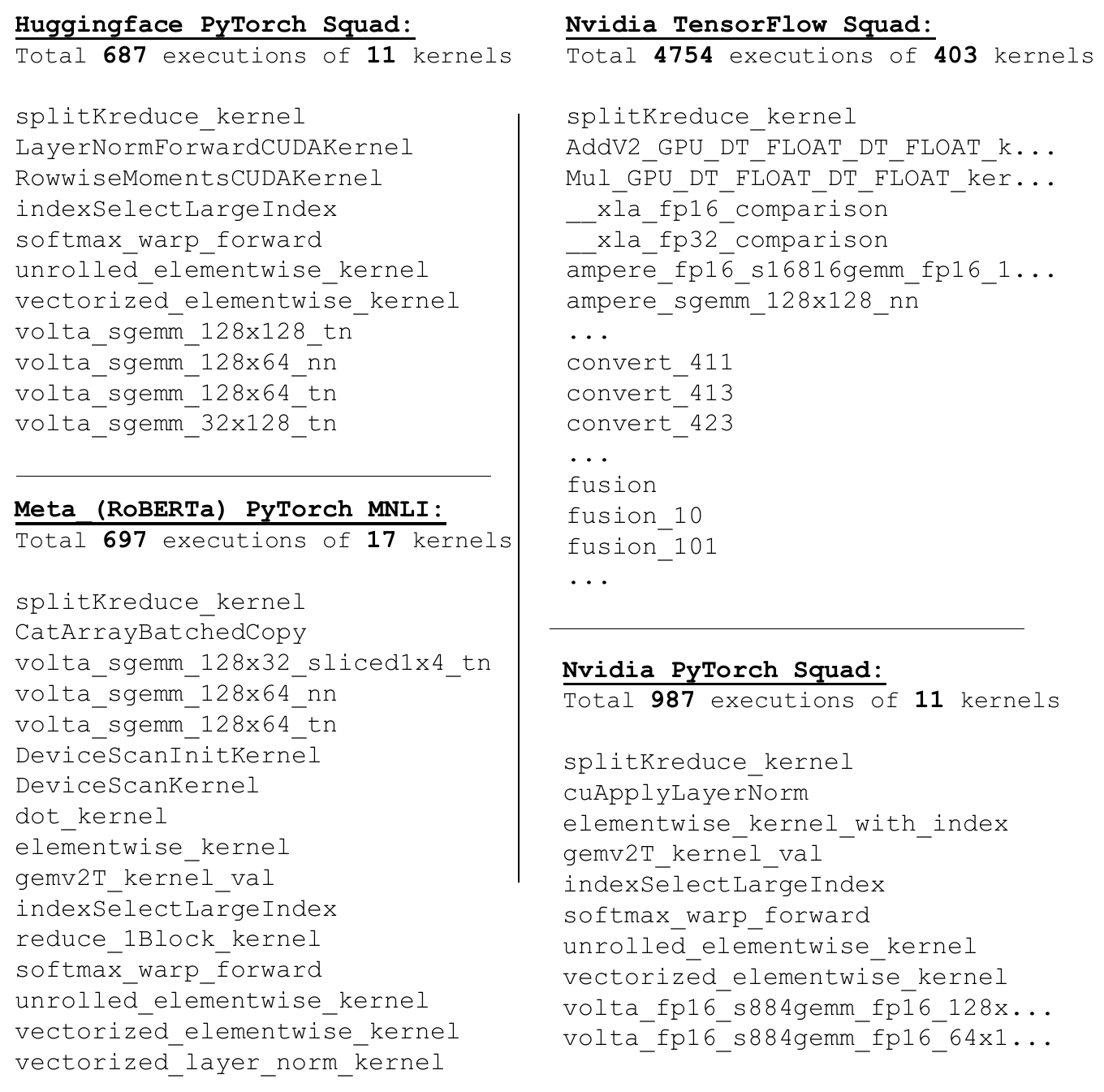}
\caption{Kernels Executed by BERT-large Models} \label{fig.kernel_bertlarge}\vspace{-10pt}
\end{figure}

The vendor signature is mainly sourced from individual vendors' preference for algorithms and frameworks, which leads to different GPU kernel selections. Figure~\ref{fig.kernel_bertlarge} shows the list of kernels executed by BERT-large models of different vendors and frameworks. Though the same model architecture was executed, only one kernel (\texttt{splitKreduce$\_$Kernel}) was commonly used by all the listed four models. From the kernel usages, we found that the most notable differences appear between different frameworks. TensorFlow models run 5$\times$ $\sim$ 8$\times$ more kernel executions and use almost 40$\times$ more unique kernels than PyTorch models. Also, TensorFlow models tend to use their GPU backends (e.g., \texttt{Mul$\_$GPU$\_$DT$\_$FLOAT$\_$DT$\_$Float}, \texttt{convert}, \texttt{fusion}, etc.), while PyTorch models use more GPU library functions (e.g., \texttt{volta$\_$sgemm}). %For the PyTorch models, across different vendors, there are some common kernels mostly GPU library functions. 
Vendor-specific kernel preferences are also observed. Though different frameworks were used, NVIDIA models are commonly optimized to leverage their Tensor Core by running functions using half-precision (fp16) data types. %Such a device-specific optimizations lead to almost 10$\times$ faster inference time than the other vendors' execution times, as can be seen in X-axis of Figure~\ref{fig.vendorvari}. 
On the other hand, Meta models tend to run many short kernels such as reduction operations (e.g., \texttt{reduce$\_$1Block$\_$} \texttt{kernel}, \texttt{DeviceScanKernel}, etc.) and hence the statistics have crowded kernel executions on the bottom of the graph. %as shown in the right bottom graph of Figure~\ref{fig.vendorvari}. 

\begin{figure}\vspace{-10pt}
\centering
  \includegraphics[width=0.33\textwidth]{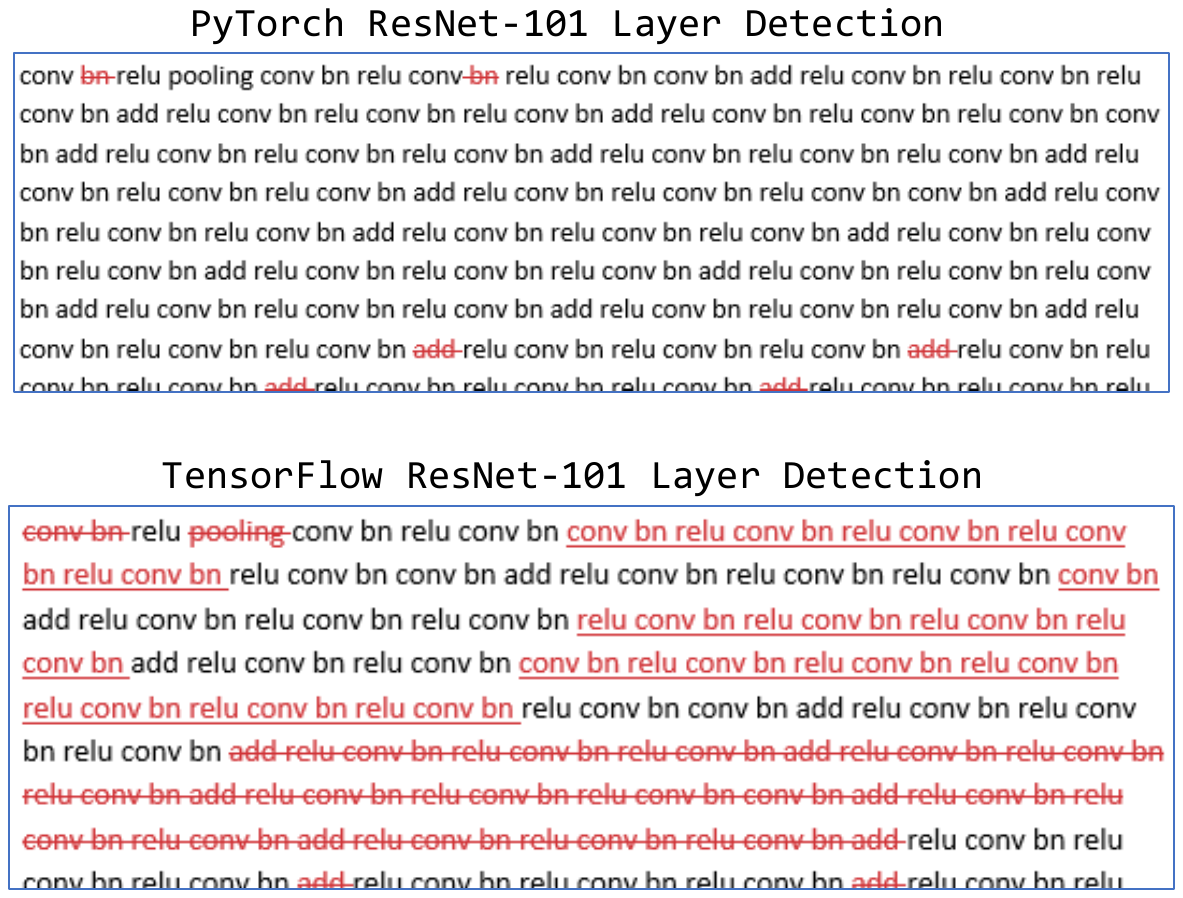}
  \caption{Impact of Framework for CNN Model Extraction Attack Accuracy: layer detection errors are colored red}
\label{fig.deepsniffer}\vspace{-10pt}
\end{figure}

The vendor signatures do not only lead to kernel execution time pattern disparity. All the related architecture hints that are commonly leveraged by many model extraction attacks %(e.g., cache misses, DRAM accesses, number of instructions, etc.) 
become different and hence the attack success rate can be significantly reduced. As most of architecture-hint-based model extraction attacks have targeted CNN models, we %Such a vendor signature is not only found in BERT models. Though the target model is different, we 
tested a state-of-the-art CNN model extraction attack~\cite{deepsniffer_asplos20}, which is designed for PyTorch models (though the authors did not discuss framework impacts), with statistics of various models that use different frameworks and released by different vendors. Figure~\ref{fig.deepsniffer} shows layer detection results when the attack model is tested with ResNet-101 models in PyTorch and TensorFlow versions. The red-colored layers are mis-predictions. There is notably higher error rate when using different framework even when the models use the same CNN architecture. Table~\ref{tab:kernel-resnet50} shows the layer prediction errors when testing with various models. LER means how many layers are incorrectly predicted per layer~\cite{deepsniffer_asplos20}. Thus, LER over 1 means that the prediction results can't be considered as having meaningful results. 
From this result, we also observed that even when using the same framework, if the model is released by different vendors (see the result of NVIDIA PyTorch in the table) and hence uses different GPU kernels, the prediction accuracy drops significantly. 

\begin{table}[h]
\centering
\caption{Impact of Framework and Vendor Optimizations for CNN Layer Detection Accuracy}\label{tab:kernel-resnet50}
\scalebox{0.85}{
\begin{tabular}{llll}
\hline
\multicolumn{1}{c}{Models}   & \multicolumn{1}{c}{\begin{tabular}[c]{@{}c@{}}Error Rate \\ (LER)\end{tabular}} & \multicolumn{1}{c}{\begin{tabular}[c]{@{}c@{}}Kernel Seq. \\ Length\end{tabular}} & \multicolumn{1}{c}{\begin{tabular}[c]{@{}c@{}}\# of Unique \\ Kernels\end{tabular}} \\ \hline
DeepSniffer Original Results~\cite{deepsniffer_asplos20} & 0.091                                                                           & 222                                                                               & 16                                                                                  \\
DeepSniffer Pytorch Model~\cite{deepsniffer_resnet}    & 0.567                                                                           & 256                                                                               & 16                                                                                  \\
Nvidia PyTorch Model~\cite{nvidia_resnet}         & 2.628                                                                           & 1235                                                                              & 38                                                                                  \\
Google Tensorflow Model~\cite{tensorflow_resnet-50}      & 6.274                                                                           & 3399                                                                              & 50                                                                                  \\
Amazon Mxnet Model~\cite{mxnet_resnet}           & 6.768                                                                           & 2652                                                                              & 59                                                                                  \\\hline
\end{tabular}}\vspace{-5pt}
\end{table}

To our best knowledge, there has not been a study that considers the impact of vendor and framework for model extraction attacks. As it is hard for an adversary to know what kind of software stack the victim black-box model uses, the attacks have higher noise than the earlier studies reported. In our study, we show how the vendor signatures can be incorporated to improve attack success rate by identifying the pre-trained model used by the black-box fine-tuned model. As noted earlier, once the baseline pre-trained model is located, the weights can be easily estimated for a whole model.

\vspace{-10pt}

\section{Threat Model}\label{sec:threat_model}

Unlike existing studies that either recover model architecture or weight values for mostly CNN-based solutions, we show that both model architecture and weights can be recovered for Transformer models. Due to the different information granularity, where architecture extraction requires individual kernel execution pattern whereas weight extraction needs bit-level information, we assume different scoped threat models for each target information.

\textbf{Adversary Identity: } Figure~\ref{fig.overview} illustrates the overall architecture of our threat model. We assume that an adversary is either an end-user that has a physical access to the device that the victim model is deployed or an admin of a cloud server where the victim model is running. The admin attacker may or may not have physical access to the server but he/she can at least run GPU performance/memory profiler that can measure various architectural characteristics. %To identify pre-trained model that the victim model is fine-tuned with, t
The adversary is assumed to have a large pool of pre-trained models that are at least publicly accessible. With the model pool, the adversary is assumed to have good knowledge about various vendor and framework signatures at least that can be observed from the target system that the victim model is deployed.

\textbf{Model Architecture Extraction: }Our architecture extraction model is a blackbox attack, where adversaries do not have any information about the model architecture. To identify the model architecture (the boundary, number, size, and algorithm of encoders that determine the victim model), the adversary measures GPU kernel execution time. If the adversary has a physical access to the target machine, he/she can use electromagnetic (EM) side-channel attack on interconnects between CPU and GPU~\cite{em_sidechannel}. If the adversary is an admin, GPU performance profiler provides sufficient information about individual kernel execution times. These side-channels are accessible as many earlier studies assumed~\cite{deepsniffer_asplos20, hoda_ccs18, mengjia_security20}.

\textbf{Weight Extraction: }The proposed weight extraction is a graybox attack, where the adversary is aware of the victim model architecture through the above model architecture extraction method. We explore two methods to extract weights: selective weight checking and heuristic testing-based weight estimation. For the selective weight checking, the adversary can acquire weight file address range through EM side-channel attack or memory profiling and check the weight values through either row hammering or bus probing, like earlier studies assumed~\cite{luo_gpusidechannel, deepsniffer_asplos20, deepsteal, mengjia_security20, hoda_ccs18}. For the heuristic testing, the attacker estimates the weights through fine-tuning of a pre-trained model with datasets that are related to the victim model's task. Thus, the only capability that the attacker additionally has is to run the victim model with various inputs and check the prediction outputs, which requires general user-level capability. 

\begin{comment}

\subsection{End-to-End Attack Scenario}

\hl{
add an example scenario that an attacker 1) collects various pre-trained models, 2) measures arch hints from a black-box fine-tuned model, 3) identify model architecture and vendor, 4) and extract weights based on the information retrieved from step 3 and pre-trained model
}
\end{comment}

\section{Model Extraction}
Inspired by the observations, we propose a new model extraction attack that exploits weight-value similarity and vendor \& framework signatures. We use BERT models as the main attack targets in this paper, though the same methods can be used for any models that follow similar architectures.

\subsection{Model Architecture Extraction}\label{sec:arch_extraction}
We examined various architecture hints such as DRAM reads and writes, cache misses and accesses, kernel execution time, the total number of kernels invoked during the inference, and so on for various fine-tuned models. We found that the time-series kernel execution time reveals sufficient model architecture information for the target BERT models. This is because BERT models use fairly regular architectures unlike other models such as CNNs. A CNN model uses different types of layers (though mostly convolution, pooling, and fully-connected) without strict ordering, and the individual layers use different weight size and number of neurons. Therefore, a combination of multiple arch hints is needed to identify layer boundaries and types of layers. %For example, DeepSniffer~\cite{deepsniffer_asplos20} uses cache miss ratio and kernel execution times to identify layer boundaries and types. It also tracks memory addresses to figure out inter-layer dependencies with which the network architecture can be identified. 
However, Transformer models including BERT run identically-shaped encoders repeatedly and hence architecture hints also show repetitive patterns. Therefore, a group of repeated measurements can be considered as one encoder. The encoders of any BERT models use the same internal layers as described in Section~\ref{sec.motivation} while the volume of computations may vary. Such encoder size can be also detected with kernel execution time because larger encoders consume longer execution time, which we will show the actual statistics soon. Therefore, we use a time-series kernel execution time as the only side channel for our attack model. 

%Note that kernel execution boundary and length can be detected without detailed profilers or performance counters that most of earlier attack models assumed to exploit~\cite{hoda_ccs18, mengjia_security20, deepsniffer_asplos20}. As a response to these attacks, GPU vendors recently added restrictions that only root privileged users can access performance counters~\cite{nv_perf}. However, performance counter access restriction cannot be a fundamental solution because the system admin may be an adversary if the model is executed on a server systems and other types of physical side-channels (e.g., power/EM probing~\cite{luo_gpusidechannel}) are possible if the model is on a mobile device. 
We assume that the kernel execution time is measured through EM side-channel attack on interconnects or a performance profiler, as assumed in the Threat model in Section~\ref{sec:threat_model}. Note that kernel execution time is less vulnerable to noise because a probing frequency that is several hundred nano-second- to mili-second-scale is sufficient to detect individual kernels and kernel boundaries do not need intra-GPU chip probing; bus activities between CPU and GPU are sufficient. Our noise sensitivity study shows robust model architecture prediction accuracy with various noise levels. Details can be found in Section~\ref{sec:noise}. %We show that attacks are still possible without detailed architecture hints in this study. %and hence blocking performance profiler accesses is not a fundamental solution for defense. 
%Instead, we use other methods such as power and EM probing to detect the activities between CPU and GPU~\cite{luo_gpusidechannel}. Though some earlier studies show that it is hard to track individual GPU threads' activities due to high parallelism, the kernel boundary only needs a detection in the level of several milliseconds or tens of nanoseconds that also involve bus transactions with CPUs, thereby easier and more accurate. Also, %It only adds cumbersome to many benign researchers to test their applications with root privileges.  %Also, the main differences among BERT variants such as BERT-base and BERT-large are the size and the number of encoders. Therefore, different BERT models can be identified by counting the number of repeated patterns.

\subsubsection{Recovering Encoder Boundary \& Count}

\begin{figure}[t]
\centering
  \includegraphics[width=0.5\textwidth]{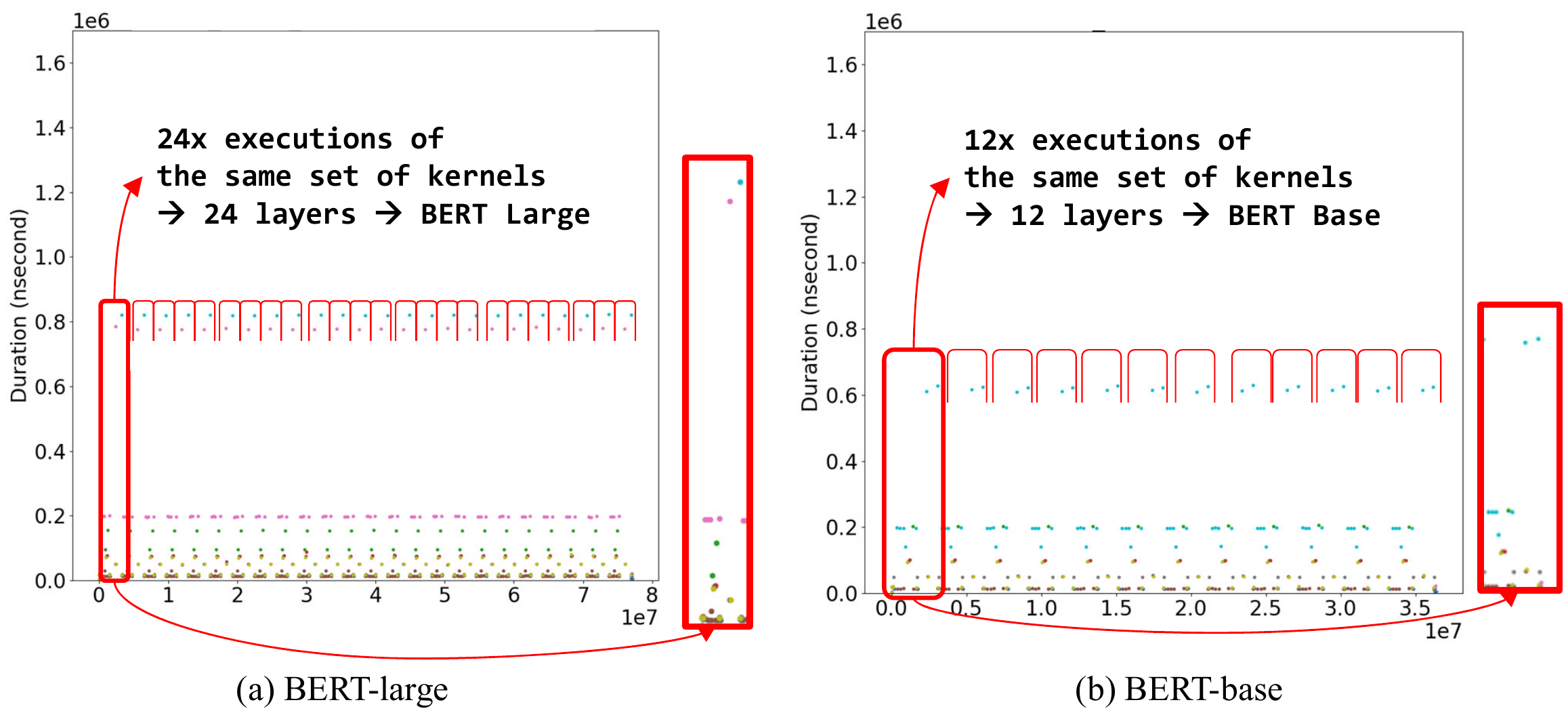}
  \caption{Encoder Boundary Identification from Time-series Kernel Execution Times: Each dot reflects the execution time of one GPU kernel}
\label{fig.layerboundary}\vspace{-15pt}
\end{figure}

\begin{comment}

\begin{figure}[t]
\centering
  \subfloat[][BERT-base ]{\includegraphics[width=0.3\textwidth]{figures/duration_layerboundary_base.pdf}}\\
  \subfloat[][BERT-large]{\includegraphics[width=0.3\textwidth]{figures/duration_layerboundary_large.pdf}}
  \caption{Encoder Boundary Identification from Time-series Kernel Execution Times: Each dot reflects the execution time of one GPU kernel}
\label{fig.layerboundary}\vspace{-15pt}
\end{figure}
\end{comment}

Figure~\ref{fig.layerboundary} shows layer identification examples for BERT-base and BERT-large PyTorch models. In each kernel execution time graph, it is easy to observe that a unit of multiple kernels is repeatedly executed. The box on each graph is the unit of kernel group. The zoom-in version is next to each graph. By partitioning the graph vertically for each kernel-group unit, BERT-base has 12 repeated patterns and BERT-large has 24 of them, which are matching with the number of encoders of each model architecture. In other words, the unit of kernel group is corresponding to one encoder execution. Without identifying individual intra-encoder layers and dependencies across encoders, the boundary of encoders and the number of them can be easily recovered. 

\begin{figure}[t]
\centering
  \includegraphics[width=0.5\textwidth]{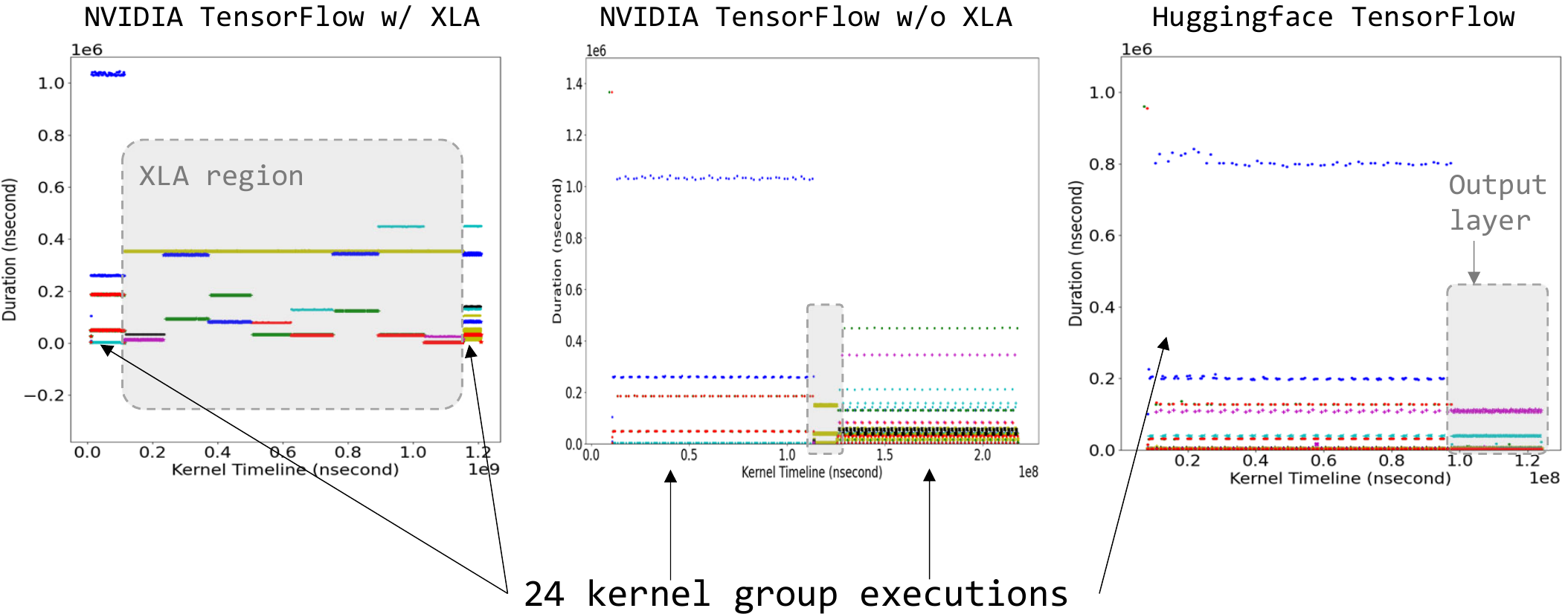}
  \caption{TensorFlow Model Execution Patterns with/wo Optimizations}
\label{fig.tf_berglarge}\vspace{-15pt}
\end{figure}

\begin{figure}[h]
\centering
  \includegraphics[width=0.4\textwidth]{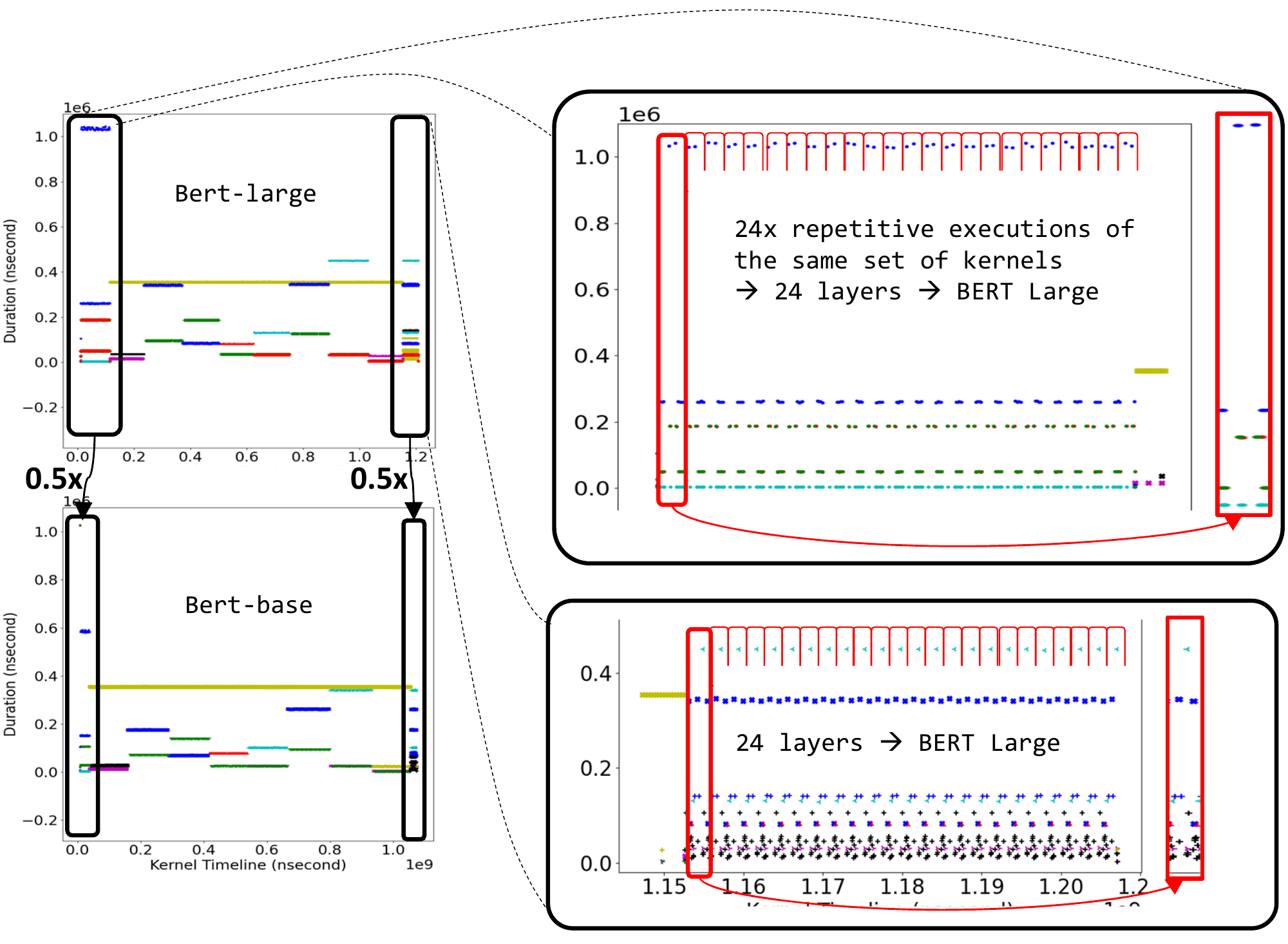}
  \caption{Encoder Boundary Identification for TensorFlow Models with XLA}
\label{fig.tf_layerboundary}\vspace{-10pt}
\end{figure}

Encoder boundary identification of PyTorch models are straightforward. However, as can be seen in Figure~\ref{fig.vendorvari}, NVIDIA TensorFlow model's statistics show notably different patterns. From digging into the script, we found that NVIDIA TensorFlow model by default uses XLA optimization~\cite{xla}. After disabling the optimization, the TensorFlow model execution also showed clear encoder boundaries. Figure~\ref{fig.tf_berglarge} shows three execution patterns of TensorFlow models. Left two graphs show NVIDIA TensorFlow models with and without XLA optimizations and the right-most graph is Huggingface TensorFlow model that does not use any optimization. It turns out XLA optimization runs compiler optimization operations in the middle of inference, as marked as XLA region. The executions for encoders are in the beginning and ending of the inference (outside the XLA regions). In both encoder regions, we were able to find 24 repetitive kernel group executions, as plotted in Figure~\ref{fig.tf_layerboundary}. %Likely, when disabling XLA, the first and the second half of the execution both consist of 24 kernel group executions. 
Huggingface model had a slight difference, where there is only one region that has 24 group executions followed by a long last layer execution. Therefore, Huggingface TensorFlow model architecture can be identified similar to PyTorch model detection method. 

For the NVIDIA TensorFlow models, especially when XLA option is enabled, the vast majority of execution is spent for XLA computations as shown in Figure~\ref{fig.tf_layerboundary}. The length and statistics of XLA region are similar in different BERT architectures. Therefore, the statistics cannot be directly used for encoder detection. Instead, we propose to develop a model that identifies the encoder regions. Within the encoder regions (highlighted with black boxes in the beginning and ending of each model graphs), the encoder detection method used for PyTorch models can be reused. The red lines show clear encoder boundaries in the zoom-in graphs on the right side of the Figure~\ref{fig.tf_layerboundary}. XLA is a TensorFlow-specific optimization. Similarly, there are other vendor-specific and framework-specific optimizations that add noise for model extraction. Therefore, one method can't be used for all. However, as far as there is any consistency found from similar models, the attacker can always make an alternative solution, just like we did for XLA optimization cases; we were able to locate the encoder region because XLA-enabled BERT-large and BERT-base models show similar patterns.

%Therefore, the same pattern recognition can't be directly applied. From investigation on TensorFlow scripts and the kernel execution time statistics of different-size BERT models, we observed that TensorFlow models run various common and memory boundary checking kernels (e.g., redzone\_checker) in the middle of the inference and hence it is hard to distinguish individual encoder boundaries. Instead, there are regions of execution especially in the beginning and ending of the inference that run kernels proportional to the number of encoders. Figure~\ref{fig.tf_layerboundary} shows the kernel execution time graphs of BERT-large and BERT-base of TensorFlow models. While the overall kernel execution patterns do not show clear encoder boundaries (e.g., some kernels such as redzone\_checker and ampere\_sgemm\_128x128\_nn are executed with the same frequency in both models), the highlighted (with black boxes) regions have groups of kernels that are executed as many times as the number of encoders. The zoom-in graphs on the right-hand side show repeated kernel groups for both regions of BERT-large. As can be seen, there are clear 24 repeated executions of the same group of kernels. Likely in BERT-base, similar kernel groups are executed 12 times. Therefore, with some pre-processing to identify the regions of execution that have hints for model architecture, the number of encoders can be also recognized from TensorFlow models with basically similar methods used for PyTorch models.

\subsubsection{Recovering Encoder Size}

Though the number of encoders provide enough information to distinguish BERT architecture, we also observed that encoder size is detectable from the kernel execution time graph. Note that BERT-base has 12 encoders where each encoder contains 12 attention heads with weight dimensions as 768$\times$768, while BERT-large has 24 encoders where each encoder includes 16 attention heads with weight dimensions as 1024$\times$1024. Therefore, an encoder of a BERT-large model needs around 70\% more computations than that of BERT-base model. In all tested models, we consistently observed that the longest kernel's execution time of BERT-large models is at least 20\% longer than that of BERT-base models. For example, in Figure~\ref{fig.layerboundary}, a BERT-base model's peak kernel execution time is around 0.6 $\times$ 1e6 ns (= 0.6 ms) while the same vendor's BERT-large model's is around 0.8 ms. In Tensorflow models, the execution regions that have architecture hint have notable peak execution time differences as can be seen in Figure~\ref{fig.tf_layerboundary}; BERT-base's peak kernel execution time is around 0.43 ms while BERT-large's is over 1 ms. With the profiled typical peak kernel execution time, it is possible to estimate the encoder size.

\subsubsection{Recovering Intra-Encoder Architecture}

Though the size and number of encoders vary, all BERT model variants use the same encoder architecture as described in Section~\ref{sec.motivation}. Each encoder consists of a self-attention layer and a feed-forward layer. %In the self-attention layer, three core vectors namely Key, Value, and Query, are generated and used for calculating an attention score. 
If intra-encoder layers and algorithms are the targets to leak (in case the victim model makes some optimizations on encoders), we observed that some post-processing on the kernel execution time statistics enables layer and algorithm detection. 

\begin{figure}[t]
\centering
  \subfloat[][BERT-base Per-Layer Stats]{\includegraphics[width=0.25\textwidth]{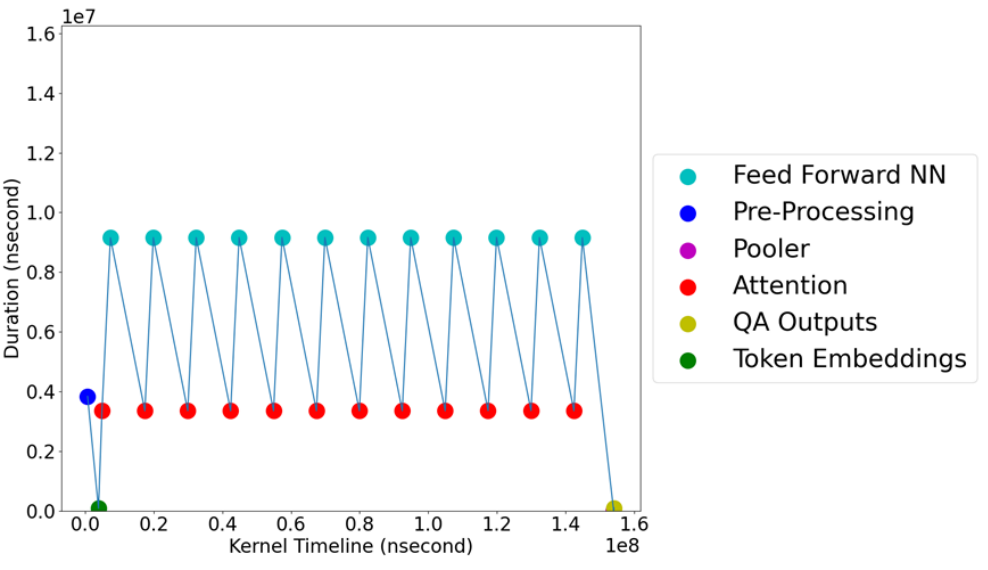}} 
  \subfloat[][BERT-base Per-Algorithm Stats ]{\includegraphics[width=0.25\textwidth]{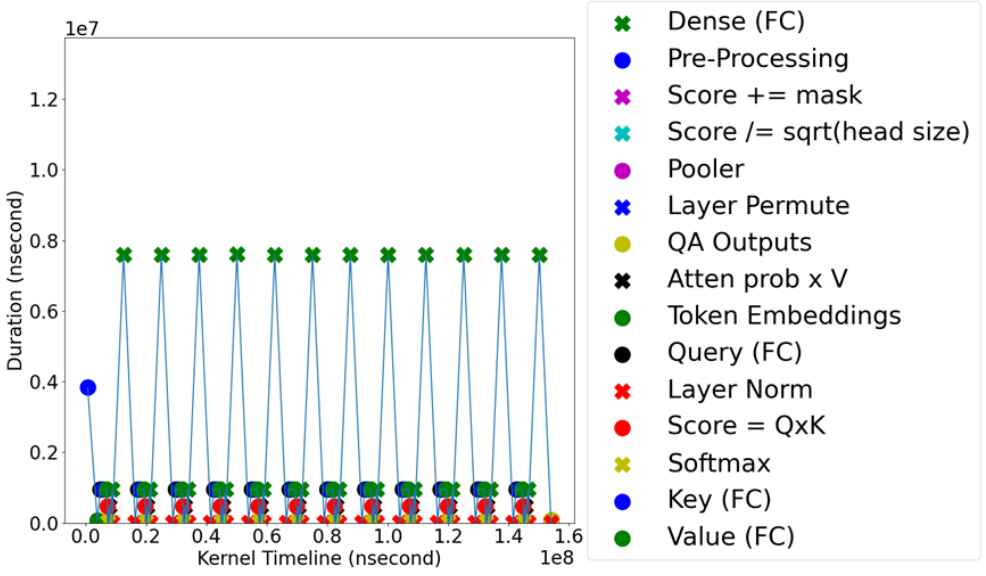}}\\
  \subfloat[][BERT-large Per-Layer Stats ]{\includegraphics[width=0.25\textwidth]{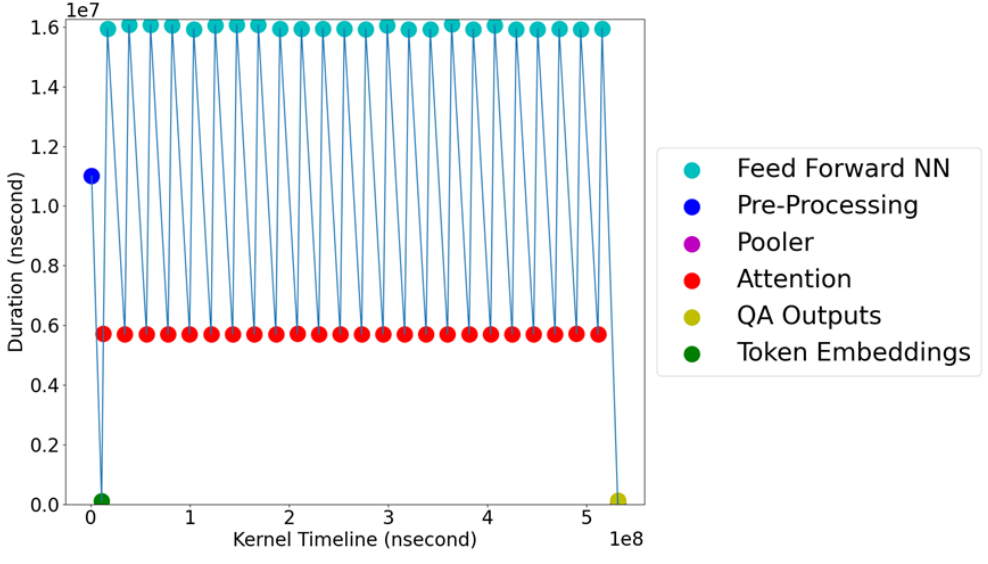}}  
  \subfloat[][BERT-large Per-Algorithm Stats]{\includegraphics[width=0.25\textwidth]{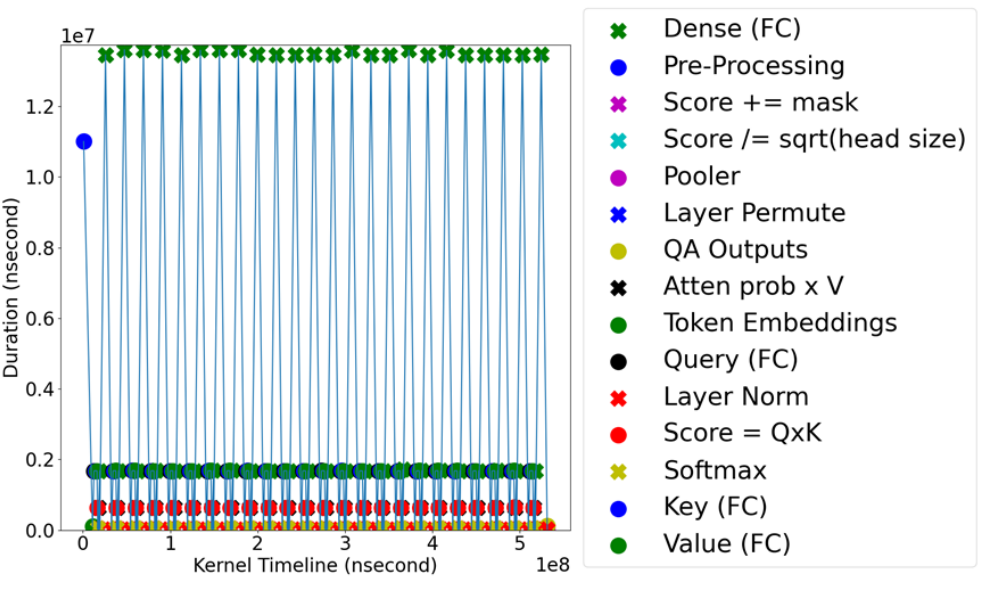}}
  \caption{Execution Times of Layers and Algorithms of BERT PyTorch SQuAD Models Released by NVIDIA}
  \label{fig.layer-fine-base}\vspace{-15pt}
\end{figure}

Figure~\ref{fig.layer-fine-base} shows example per-layer and algorithm execution times of BERT-large and BERT-base models. In both models, we were able to find that feed-forward layers take over 2$\times$ longer execution times than attention layers. For each attention layer, the Key, Value, and Query vector calculations equally consume the most time. The attention score calculation is the second most time-consuming operation. Softmax and layer-norm operations are consistently very short. Therefore, from this layer- and algorithm-boundary analysis, adversary will be able to identify encoder architecture. For example, if there are more than three equal-length layers that consume the highest time within each attention layer, the adversary can catch that the victim model is using one more core vectors besides, Key, Value, and Query. Also, if per-layer execution time is less than typical duration, the adversary can suspect other optimizations such as head pruning. We discuss the pruned head prediction in Section~\ref{sec:quant}.

This layer- and algorithm-boundary detection needs vendor signature detection apriori because each layer typically runs multiple kernels, and different vendors and frameworks use different kind of kernels, as discussed with Figure~\ref{fig.kernel_bertlarge}. %For example, for the same layer norm operations, Huggingface models use three GPU kernels (vectorized\_elementwise\_kernel, RowwiseMomentsCUDAKernel, and LayerNormForwardCUDAKernel) while NVIDIA models use two kernels (vectorized\_elementwise\_kernel and cuApplyLayerNorm). 
We assume that the attacker already has vendor signature information through exhaustive evaluations. %on various white box models. %that the attacker already has.

\subsubsection{Automated Model Architecture Extraction}

To automate the attack process, there are two challenges: 1) the kernel group shape varies due to vendor signatures and 2) the kernel groups can be only distinguishable in the time domain, which means individual kernels' execution times as well as invocation timing information are needed to distinguish kernel group repetitions. Therefore, the automation needs to be able to 1) detect kernel groups when the groups have different size and shape and 2) process a 2-dimensional information, where each dimension indicates the kernel invocation sequence and individual kernel execution time. We found that this problem can be reduced to a pattern recognition problem for 2-dimensional inputs, which is similar to image recognition. Each kernel group is the target pattern to detect and the time-series kernel execution graph can be considered as an input image. Therefore, we propose to use a CNN model for recovering the victim model architecture. CNN is especially effective for this problem because it can recognize various patterns (divers kernel group shapes) for individual object class (kernel group that corresponds to an encoder). For example, our encoder detection is similar to a vehicle recognition where vehicles with various types and colors can be equally classified to a vehicle. 

We explored various architectures and finalized the CNN model with seven hidden layers: conv (in: 3, out: 6, kernel: 5 $\times$ 5), pool (kernel: 8 $\times$ 8, stride: 8), conv (in: 6, out: 16, kernel: 5 $\times$ 5), pool (kernel: 8 $\times$ 8, stride: 8), fc (in: 3600, out: 120), fc (in: 120, out: 84), fc (in: 84). The model is trained with time-series kernel execution time graphs. We converted each graph to a gray-scale image where each pixel is colored with either white or black. Each pixel on a x-y tuple as (kernel invocation time, kernel execution time) has black color. We made all graph images uniformly have 1024 $\times$ 1024 pixels.  The converted graph images were collected while running the inferences of 10 BERT models listed in Table~\ref{tab.models} (except for CNN and GLUE benchmark). %544 images were used for training and 136 images as testing. 
The ground truth of each prediction is the number of encoders in the input graph. %The CNN detects the patterns of encoders and predicts how many encoders are in each time-series kernel execution graph. 
The number of encoders is the most important parameter that represents a certain BERT model architecture as the widely used BERT-base and BERT-large have static number of encoders. %For example, if there are 12 encoders, we can estimate that the victim model is BERT-base and if there are 24 encoders, the model is likely to be BERT-large. 
Though BERT-base and BERT-large are the two widely used architectures, we also trained the model by changing the number of encoders gradually such that any victim models that are optimized to use arbitrary number of encoders can be also recovered.
\vspace{-2pt}
\subsection{Weight Extraction}
\vspace{-2pt}
\subsubsection{Recovering Vendor \& Framework}\label{sec:vendor_extraction}
With a simple CNN model, BERT architectures can be detected. To extract weights, we also need to detect vendor and framework. For this purpose, we propose to develop two more CNN models each predicts vendor and framework of the pre-trained model used by the victim model. % is fine-tuned with. 
This is again similar to vehicle detection but with a finer-grained prediction that detects types of vehicles separately, such as sedan, truck, suv, and so on. For the training dataset collected for the encoder detection, we re-labeled them to individual vendor and framework.  The same CNN architecture worked accurately. 
We assume that an adversary has a large pool of pre-trained models that are publicly released. With the predicted model architecture, vendor, and framework, the attacker can search the pool of pre-trained models. Once a matching pre-trained model is found, the weight values of the pre-trained model is regarded to be almost similar to the victim fine-tuned model's, as we observed in Section~\ref{sec.motivation}.

\textbf{Model Variants Detection:} Some vendors release several variant models for the same BERT architecture and framework, namely \textit{cased}, \textit{uncased}, \textit{whole word masked} and \textit{sub-word masked}. We found that these variant models show almost alike execution patterns as they follow the same architecture and use similar GPU kernels. For these models, we use special input testing-based recognition by exploiting the dedicated purposes that each model is designed for. For example, a cased model recognizes different meanings when a word is written in upper case and lower case, while an uncased model cannot recognize the differences. %Therefore, it is possible to recognize these variants through testings with special inputs. For example, 
Thus, when a context is provided with Apple as a company name and apple as a fruit name used together, the cased model and the uncased model predict differently. Once the variant model is detected, the weight values can be also extracted as described earlier. We tested all four variants of Huggingface BERT-large model and found that the fine-tuned model weights are different with each other with up to 0.25 value distance, while each fine-tuned model and its baseline pre-trained variant model show very high similarity (< 0.002 value distance). Therefore, once vendor, framework, and variant information are detected, almost similar weights can be recovered from the corresponding pre-trained model.
%\hyeran{add discussions about weight value similarity across models fine-tuned for different tasks that have the most significant differences only in the last layer}

\subsubsection{Recovering Actual Weights}
Though our pre-trained model identification recovers all encoder layers' weight values with very small (< 0.002) differences, the weight value gap can be further reduced by applying two methods: 1) selective weight checking and 2) heuristic testings. The last classification layer weights can be recovered with the combination of these two methods. % With the pre-trained model found from the vendor and framework detection, the attacker can further extract actual weight values by using heuristic testings 

\begin{figure}[h]
\centering
\includegraphics[width=0.45\textwidth]{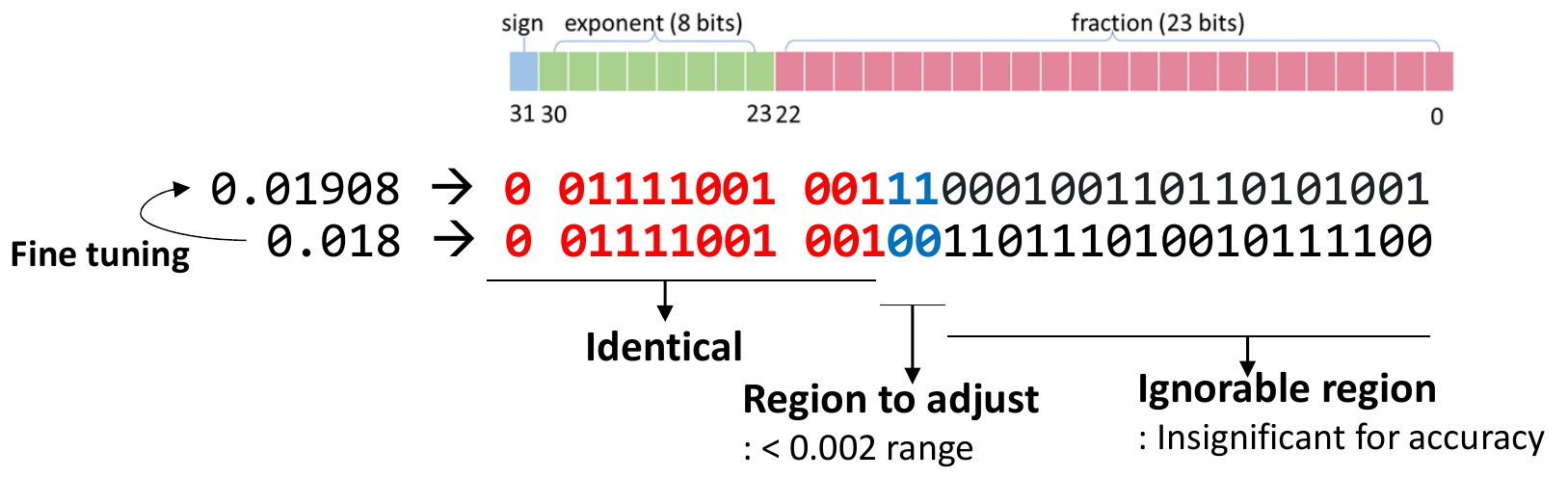}
\caption{Selective Value Checking}
  \label{fig.hammer}\vspace{-10pt}
\end{figure} 

\textbf{Selective Weight Checking with Weight Extraction Pruning: }
Selective value checking leverages existing solutions for individual weight value extraction. Some studies used bus probing and some other studies leveraged rowhammer attacks on the victim model (fine-tuned model in our case)~\cite{deepsteal}. %follows similar approach used by HammerLeak~\cite{deepsteal}. HammerLeak leverages rowhammer attack to reveal individual bit values of CNN model weights. 
Those studies showed that it is possible to reveal weight values for short data types (e.g., integer values) for CNN models. %They successfully recovered short integer weight values up to 8 bits. 
However, individual weight value recovery requires hundreds of hammering and bus probing and hence it is not desirable to be used for longer data types such as float16 or float32. But, our weight value similarity can effectively reduce the scope of checking. As pre-trained model and its fine-tuned model has < 0.002 value difference, sign and exponent fields rarely change. %when a pre-trained model is fine-tuned. 
From our experiments, we observed that average of around 99\% weights keep their sign when fine-tuned. Only the fraction part has differences mostly. Out of them, we found that only one to two bits that correspond to 0.002 range need to be checked. Figure~\ref{fig.hammer} shows an example. When a weight value 0.018 is fine-tuned to 0.01908, the sign and exponent fields are identical. Also, first a few bits of fraction field are identical, as highlighted with red color. Only the two bits in blue color that make the differences between the two values may need checking. The remaining 18 bits in fraction field do not need to be checked because those make differences less than 0.001. From our experiments, we observed that a fine-tuned model provides almost similar accuracy (F1 score is slightly dropped from 87.9078 to 87.8881) when all weights are adjusted to discard values below floating-point fourth digits (e.g., 0.0013 is adjusted to 0.001). Similarly, the number of weights to check can be significantly reduced because the weights that have their absolute value below 0.001 barely have impact for final prediction accuracy and hence can be excluded from checking. 

With these observations, we propose \textit{weight extraction pruning}. The pruning is done in two phases on the fine-tuned victim model for the bits that are considered unnecessary for checking, based on the weight value in the pre-trained model found from our vendor/signature detection. First, the weights whose mapping weight's absolute value in the pre-trained model is less than 0.001 are excluded from checking. Second, for the remaining weights, two bits per weight value are only checked which are corresponding to 0.002 value ranges. In the second phase, the bit locations to check may vary depending on the exponent field value and integer part of the weight value. For example, in Figure~\ref{fig.hammer}, the most significant 4$^{th}$ and 5$^{th}$ bits of fraction field are checked because the exponent part value is 121, which means that the value is multiplied by 2$^{121-127}$ = 2$^{-6}$. This means that the value's implicit integer part is corresponding to 0.015 (=2$^{-6}$) and the fraction part's first 1, which is on the most significant 3$^{rd}$ bit is corresponding to 0.0019 (=2$^{-9}$). When adding these two values, the value becomes around 0.0169. As 0.018's potential fine-tuned value range is between 0.016 and 0.02, the following two bits (highlighted with blue color) that each corresponds to 0.00097 (=2$^{-10}$) and 0.00048 (=2$^{-11}$) together make sufficient value range that determines the fine-tuned value. This bit location finding to check can be implemented in a script and the remaining bits can be excluded from checking. With almost similar value baseline acquired from the pre-training model identification, we can make the hammering/probing-based weight-value recovery more feasible for longer data formats. Algorithm~\ref{alg:swc} is a pseudo code for float32 weights. The algorithm can be revised to support different data formats by changing float32-specific formulas such as $exponent - 127$. 

\begin{algorithm}
\caption{Selective Weight Checking with Weight Pruning: Example of float32}\label{alg:swc}
\begin{algorithmic}
\State $target\_weight \gets W_{i}$ in fine-tuned model
\State $base\_weight \gets W_{i}$ in pre-trained model
\State $abs\_base\_weight \gets abs(W_{i})$ in pre-trained model
\If{$base\_weight\ <\ 0.001$}
    \State No need to check. clone model's $W_{i} \gets$ $base\_weight$
\Else
\State $min \gets base\_weight - 0.002$
\State $max \gets base\_weight + 0.002$
\State $k \gets$ most significant non-zero bit location of fraction 
\State field in $abs\_base\_weight$
\State $int\_base \gets 2^{exponent - 127}$ of $abs\_base\_weight$
\State $j \gets$ 2
\While{$j > 0$}
\State $fr\_base \gets 2^{exponent - 127 - k}$ of $abs\_base\_weight$
\If{$min \leq (int\_base$ + $fr\_base) \leq max$} 
    \State check (hammer) $k_{th}$ bit of $target\_weight$.
    \State clone model's fraction field $k_{th}$ bit of $W_{i}$ 
    \State $\gets$ fraction field $k_{th}$ bit of $target\_weight$
\EndIf
    \State $j \gets j-1$, $k \gets k+1$
\EndWhile
\EndIf
\end{algorithmic}
\end{algorithm}
\begin{figure}[t]
\centering
  \includegraphics[width=0.35\textwidth]{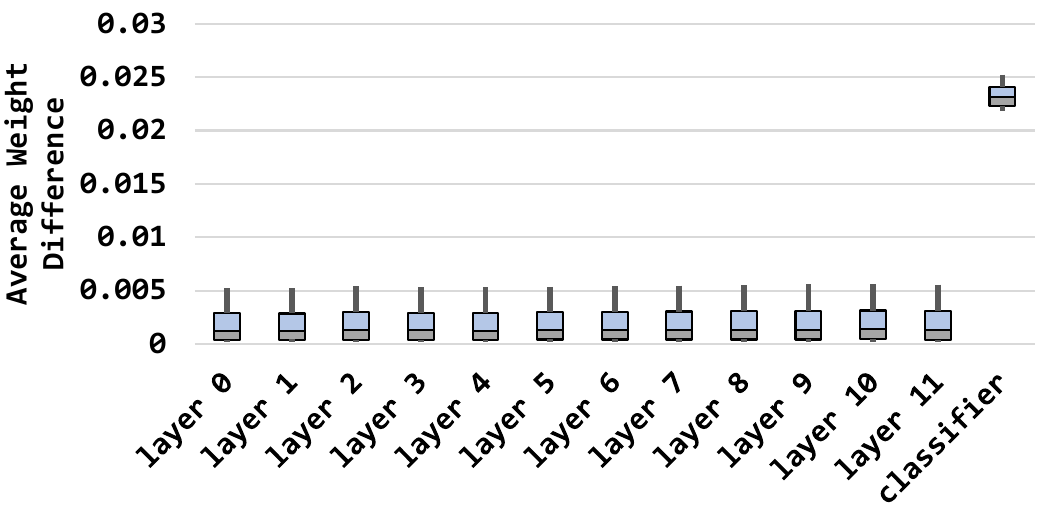}
  \caption{Average Weight Differences Among 9 BERT-base Models in GLUE benchmark}
\label{fig.weight_diff}\vspace{-15pt}
\end{figure}

\textbf{Heuristic Testing: }
When either row hammering or bus probing can not be used, heuristic testing can be also considered to estimate the actual weight values. Krishna et al.~\cite{krishna2019thieves} created functionally equivalent clone models through heuristic testing. They picked a random pre-trained BERT model, tested the victim BERT model with a list of inputs that are related to the victim model's task, and fine-tuned the random model with the outputs of the victim model to clone the victim model. They did not attempt to reveal the actual architecture of the victim model but showed that using the same BERT architecture with the victim model is more effective for cloning. Therefore, an attacker can leverage the heuristic testing over the pre-trained model that is recovered from our model architecture and vendor/signature extractions to estimate the actual weight values. In Section~\ref{sec:heuristic_result}, we apply the heuristic testing to our attack model and show the benefit of finding the actual architecture of the victim model. 

Heuristic testing can be used for reducing the scope of weight value checking for the last classifier layer. The weights of all encoder layers in BERT model are adjusted based off that of the pre-trained model during the fine-tuning. But, the last classifier layer is newly attached to be trained for the fine-tuning task. Thus, the weight value similarity might not be consistently observed. When we checked weight value distance among nine Huggingface GLUE benchmark~\cite{glue} models that are fine-tuned for different tasks from the same pre-trained model, all the encoder layers had almost zero distance while the last layer showed 10$\times$ higher differences than the other layers. Thus, the selective weight pruning might not be directly applicable for the last layer. Instead, we can consider leveraging heuristic testing. With heuristic testing, we can acquire baseline weights of the last layer and then apply selective weight checking to find the actual weight values. This works if the vendor/framework detection is successful, because similar initial weight values (e.g., random seed) are likely to be used for training the last layer.
\vspace{-5pt}
\subsubsection{Supporting Quantization and Pruning}\label{sec:quant}
We explained the weight extraction with an example of float32 data format. However, the victim model might be optimized with quantization and pruning techniques, while pre-trained models are typically released in float32 format for a highly accurate fine-tuning. The popular quantization supported by most of the frameworks are int8, uint8, int16, uint16, int32, bfloat16, and float16. For all floating-point data types, the only differences are the lengths of exponent and fraction fields. For example, compared to float32 data format, float16 uses shorter (5-bit) exponent and (10-bit) fraction fields, while bfloat16 uses the same-length (8-bit) exponent with a shorter (7-bit) fraction field. Therefore, our proposed selective value checking can be directly applied with a slight bit range adjustment. %If bfloat16 is used in the example of Figure~\ref{fig.hammer}, as bfloat16 uses the same length exponent with float32, the same bits (the blue-colored bits in the fraction field) should be checked. %Similarly, the bits to check for integer weights can be identified once the integer value range mapped to the 0.02 weight difference that is mapped to scale factor used for the quantization. 
The kernel execution time patterns provide hints for data formats. For example, according to our experiments, kernels using float32 weights have at least 2$\times$ longer execution time than float16 weights. The specific data formats used for quantization vary depending on frameworks. Thus, by leveraging the framework information extracted in Section~\ref{sec:vendor_extraction}, the corresponding pre-trained model weight file can be quantized by using the framework supported APIs before extracting actual weights. 

%\hyeran{add the quantization execution results to discuss quantization identification}
\begin{figure}[t]
\centering
  \includegraphics[width=0.5\textwidth]{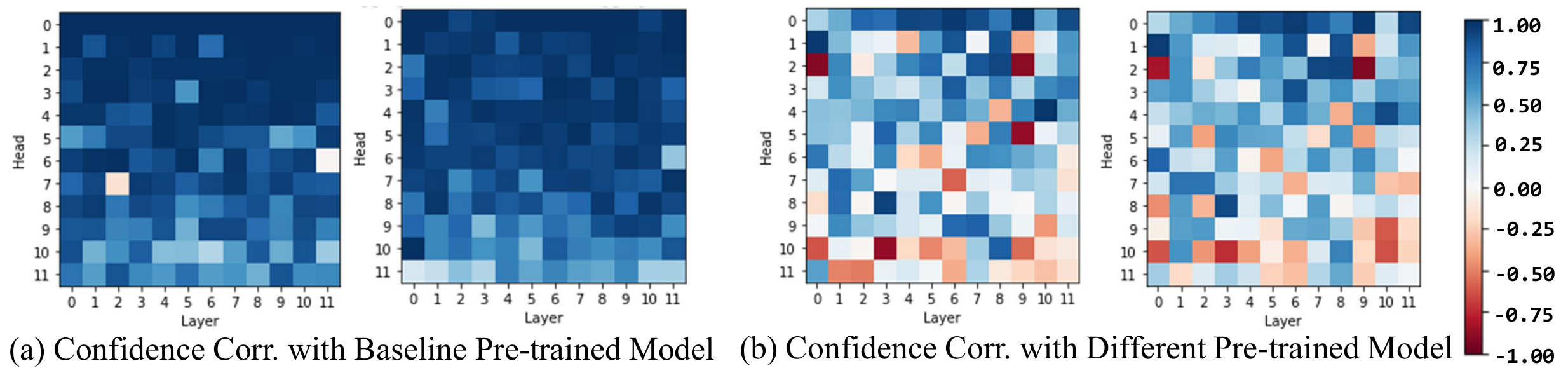}
  \caption{Pearson Correlation Coefficient of Confidence used for Head Pruning: X-axis is layer and Y-axis is head. %Whiter colors mean lower correlation and Darker Blue colors mean higher positive correlation
  }
\label{fig.head_pruning}\vspace{-15pt}
\end{figure}

The victim model might be optimized for better performance and accuracy. For example, head pruned models are known to have better accuracy and computation efficiency by removing insignificant heads~\cite{headpruning-1, headpruning-2}. The head pruning algorithm uses various metrics to identify the removable heads. One of the most popular metrics is \textit{Confidence}, which is calculated by averaging the maximum attention weights. The heads that have lower confidence than a pre-defined threshold are pruned. %This means that any two models that have very similar head confidences will have almost similar head pruning results. 
If the target victim model has some heads pruned, weight extraction will be challenging because weight locations will not be directly mapped to the unpruned pre-trained model's. For such case, we believe the attacker can acquire an important hint from the weight value similarity. Because the confidence is calculated based on the weights, if the weights are similar, head confidences are likely to be similar. With the confidence value calculated with pre-trained model's weight, the attacker can locate the pruned heads. We evaluated the confidence value similarity between a pre-trained model and its two fine-tuned models that are trained for different tasks (semantic equivalence and sentiment analysis) by feeding the same set of input queries as shown in Figure~\ref{fig.head_pruning}(a). Each cell in the matrix is a Pearson correlation coefficient between the confidence of the heads on the same location in a pre-trained model and a fine-tuned model. As the scale bar on the right indicates, darker blue cells mean high correlation while whiter cells mean no correlation. In both fine-tuned models, confidence of all heads are highly correlated with that in the pre-trained model. On the other hand, when we compared the two fine-tuned models and a different pre-trained model, the correlation dropped significantly as plotted in Figure~\ref{fig.head_pruning}(b). From this result, we believe it is challenging but possible to extract head pruning information of a victim model by leveraging the weight value similarity. %However, the weight extraction for pruned model is out of scope of our study and we would like to explore methods for the pruning extraction in our future work. }

%\subsection{Attack Noises}
%\subsubsection{Impact of Different Tasks}
%\subsubsection{Impact of Different Training Dataset}

\vspace{-2pt}

\section{Evaluation}\label{sec:experiment}

\vspace{-2pt}

\subsection{Methodology}
We examined a total of 20 models for evaluation as listed in Table~\ref{tab.models}. 10 pairs (pre-trained and fine-tuned) of BERT/ RoBERTa models are downloaded from four vendors' repositories to develop a model architecture extraction attack. For a fair comparison, we used whole-word masked and uncased models because some vendors provide only this version models. We intentionally included models from different vendors that use different deep learning frameworks to evaluate the vendor/framework signature extraction accuracy. For these 10 models, we used vendor-provided fine-tuned models rather than designing our own fine-tuned models to avoid any misinterpretation caused by our premature fine-tuning. We also used another nine fine-tuned models that are developed by using the pre-trained model and fine-tuning script of Huggingface GLUE benchmark~\cite{glue} for verifying the impact of different tasks for the weight value similarity. A ResNet-18 model is also fine-tuned based on the pre-trained model downloaded from PyTorch vision repository~\cite{torch_hub} to examine the scalability of our approach to non-Transformer model. We tested these models on a laptop that has NVIDIA GeForce GTX 1660Ti (Turing) GPU with CUDA v11.5 as well as a server that has NVIDIA V100 (Volta) GPU with CUDA v11.1. We used Python v3.8, PyTorch v1.11.0, and TensorFlow v2.8.0 for evaluations.

%For BERT question answering, the context is "My name is Harry and I grew up in Canada. I love apples." and the question is "What food does Harry like?". For GPT2, the seed is "You're my brother and ".

\begin{table}[]
\centering
\caption{20 Tested Models: The models are trained for the tasks (with datasets) in \textit{Pre-trained} and \textit{Fine-tuned} columns.}\label{tab.models}
\scalebox{0.75}{
\centering
\begin{tabular}{ccccc}
\hline
Model                       & Vendor      & Framework  & Pre-trained & Fine-tuned\\ \hline
\multirow{13}{*}{BERT-Base}  & NVIDIA~\cite{nv_bert_pt}     & PyTorch & \multirow{2}{*}{MLM/NSP} & \multirow{2}{*}{SQuAD}  \\
                            & NVIDIA~\cite{nv_bert_tf}      & TensorFlow & & \\ \cline{2-5}
                            & Huggingface~\cite{hf_bert_pt} & PyTorch & \multirow{2}{*}{MLM/NSP} & \multirow{2}{*}{SQuAD}  \\
                            & Huggingface~\cite{hf_bert_tf}      & TensorFlow &  &  \\ \cline{2-5}
                            & \multirow{9}{*}{\begin{tabular}[c]{@{}c@{}}Huggingface \\ GLUE benchmark\end{tabular}~\cite{glue}} & \multirow{9}{*}{PyTorch} & \multirow{9}{*}{MLM/NSP} & CoLA\\
                            & & &  & SST-2\\
                            & & &  & MRPC\\
                            & & &  & STS-B\\
                            & & &  & QQP\\
                            & & &  & MNLI\\
                            & & &  & QNLI\\
                            & & &  & RTE\\
                            & & &  & WNLI\\\hline
\multirow{4}{*}{BERT-Large} & \multirow{2}{*}{NVIDIA~\cite{nv_bert_pt}}      & \multirow{2}{*}{PyTorch}  & \multirow{3}{*}{MLM/NSP} & SQuAD \\
                            &                               &   &   & SST-2 \\
                            & NVIDIA~\cite{nv_bert_tf}      & TensorFlow &  & SQuAD \\ \cline{2-5}
                            & Huggingface~\cite{hf_bert_pt} & PyTorch    & \multirow{2}{*}{MLM/NSP} & \multirow{2}{*}{SQuAD} \\
                            & Huggingface~\cite{hf_bert_tf} & TensorFlow    &  &  \\ \hline
RoBERTa-Large               & Meta~\cite{mt_roberta_pt}        & PyTorch    & MLM & MNLI \\\hline 
ResNet-18                   & Meta~\cite{torch_hub}                  & PyTorch & ImageNet'12 & Hymenoptera\\\hline
\end{tabular}
}
\end{table}

\subsection{Model Architecture Extraction Accuracy}\label{sec:noise}

\begin{figure}[h]
\centering
  \includegraphics[width=0.5\textwidth]{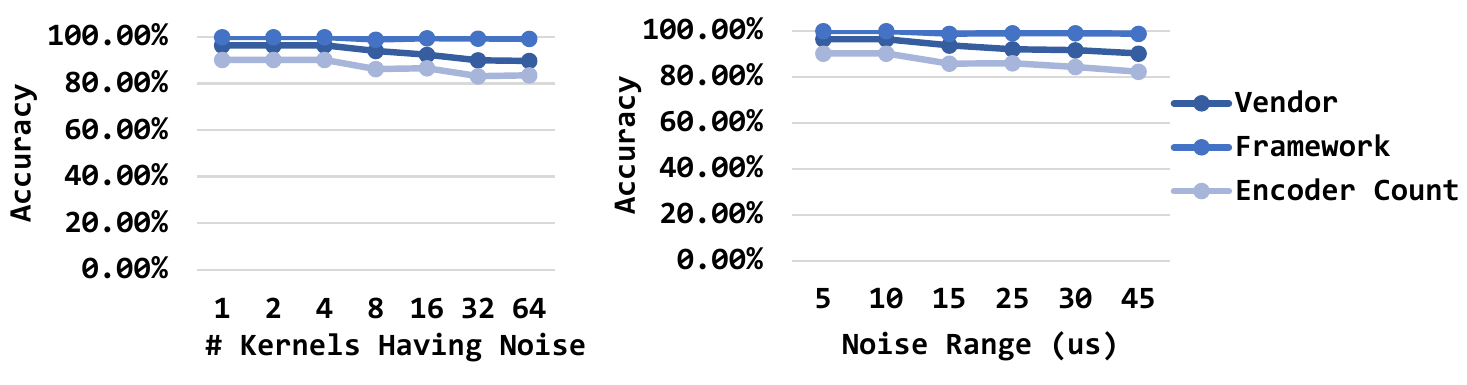}
  \caption{Extraction Accuracy with Kernel Execution Time Measurement Noise: (left) impact of kernel count having noise (right) impact of noise lengths}
\label{fig.noise}\vspace{-5pt}
\end{figure}

\begin{comment}

%\hyeran{add accuracy for vendor, number of layers, encoder size, and framework}
\begin{table}[h]
\centering
\caption{Architecture, Vendor, and Framework Prediction Per Noise Level}\label{tab.prediction}
\scalebox{0.65}{
\begin{tabular}{ccll}
\hline
Prediction                      & \multicolumn{3}{c}{Accuracy}  \\ \cline{2-4} 
\multicolumn{1}{c}{Noise Level} & \multicolumn{1}{c}{0} & 1 & 2 \\ \hline
Vendor                          & 96.30\%               &   &   \\
Framework                       & 99.85\%               &   &   \\
Encoder Count                   & 90.07\%               &   &   \\ \hline
\end{tabular}
}
\end{table}
\end{comment}

We trained the CNN-based architecture, vendor, and framework extraction attack models with 5-fold cross validation on a total of 680 profiled data. 136 data were used for testing. Thanks to the notable execution patterns, vendor, framework, and encoder count were predicted with 96.30\%, 99.85\%, and 90.07\% accuracy, respectively. 

\textbf{Noise Sensitivity : }
We evaluated the prediction accuracy by adding noise to the kernel execution time measurements. We evaluated noise sensitivity in two ways: 1) varying the scope of kernels impacted by noise 2) varying significance of noise for individual kernels. Based on the typical kernel duration, we set the noise in the first test to 20 us. We randomly selected $N$ kernels from one of the input images used for the CNN model inference and adjusted their execution time to $original\ execution\ time\ \pm 20us$, while varying $N$ from 1 to 64. For the second test, we selected 16 kernels from one of the validation images and changed their execution times to $original\ execution\ time\ \pm K us$, where $K$ is varied from 5 to 45 us. Then, the adjusted time-series kernel execution images are fed to all CNN models for vendor, framework, and architecture prediction. Figure~\ref{fig.noise} shows the accuracy results. As plotted in the figure, accuracy was dropped very slowly for both cases. This is because CNN architecture is inherently error-tolerant~\cite{li_sc17}. 

\subsection{Model Cloning Accuracy}\label{sec:heuristic_result}
We evaluated the effectiveness of heuristic testing. We used the same method as used by Krishna et al.~\cite{krishna2019thieves}, except for that the adversary can extract the victim model architecture and hence use the correct pre-trained model for heuristic testing. %However, unlike the Krishna's approach that used randomly selected BERT model as pre-trained model, we can use correctly identify the model architecture, even the pre-trained model that the victim model is likely to be fine-tuned. This adversary's capability difference makes the correct BERT model as pre-trained model 
%We used a Huggingface BERT-base PyTorch model that is fine-tuned for question-answering task as a victim model. Then, we created clone models with 1) randomly selected BERT models (no clue about victim model architecture as assumed by Krishna) and 2) the pre-trained model that is recovered by our architecture extraction attack. From the architecture extraction, we were able to identify Huggingface BERT-base PyTorch model as method 2)'s baseline. 
To compare with Krishna's approach, we also selected random pre-trained models for attack baseline models that either uses different BERT architecture or uses the correct BERT architecture but created by different vendor. %For method 1), we chose a model that uses different BERT architecture (Huggingface BERT-large pre-trained model) and another model that uses the correct BERT architecture but is released by a different vendor (NVIDIA BERT-base pre-trained model). 
Then, we fed 87599 inputs to the victim model and collected the corresponding outputs. The attack baseline models were fine-tuned with 87589 input and output pairs collected from the victim model. Note that the output here is not the dataset's ground truth; it is victim model's outputs because the goal of this attack is to create a clone of the victim model. Two epochs were used with 3e-05 as learning rate. Then, we tested the clone models with the remaining 10 inputs and evaluated the accuracy and prediction similarities with the victim model. %We collected 87599 input and output pairs with the victim model and trained clone models with these data. 
%Out of 87599 input-output pairs, 10 pairs were excluded from training and used for testing.  %The cloned models showed almost identical predictions with the victim model. However, 
The model that used correct pre-trained model showed 100\% identical outputs with the victim model while the models that used random pre-trained model derived 90\% similarity with the victim model. %Interestingly, for one of the testing data that the victim model predicted incorrectly, our recovered model predicted exactly the same (incorrect) output while the other two models predicted correct answers (ground truth of the dataset). 
The clone models that used the same BERT architecture with the victim model derived similar F1 scores with the victim model (87.9078, 88.3654, and 88.026), while the clone model that used a larger (BERT-large) pre-trained model showed even better F1 score, which is 89.97366. This is because BERT-large models can extract more complex relations better than BERT-base models due to larger and more encoders. However, as the attack goal is to make a clone of the victim model, not creating an outperforming model, our recovered pre-trained model showed a better capability to copy the victim model than randomly selected models.

\subsection{Weight Extraction Efficiency}
%Without knowing the baseline weight values and the typical weight value scales, weight extraction attacks need enormous amount of hammering or bus probing because every single bits of all weights need to be checked individually. 

%With our novel observation about value similarity between the pre-trained model and fine-tuned model, weight probing/hammering can be significantly pruned. We examined how much checking can be reduced by checking a randomly selected model's pre-trained and its fine-tuned weight values. 

\begin{figure}[h]
\centering
  \includegraphics[width=0.4\textwidth]{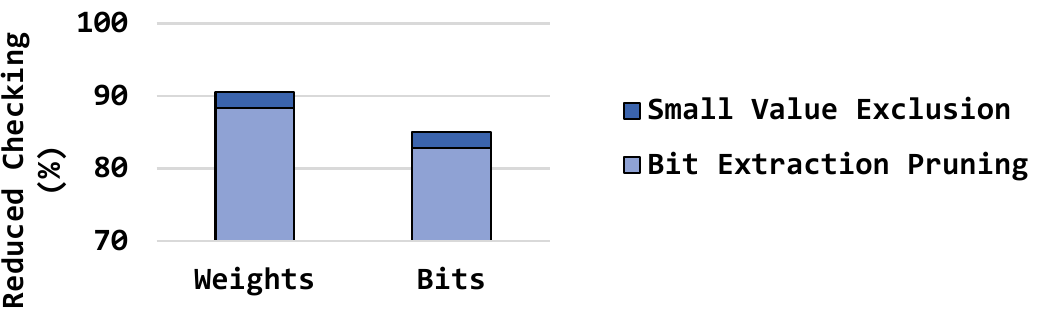}
  \caption{Reduced Weight Value Checking w/o Errors}
\label{fig.pruning_efficiency}\vspace{-10pt}
\end{figure}

To understand the efficiency of weight extraction pruning, we examined weight values of a randomly selected pre-trained model and its fine-tuned model and measured the total number of bits that need to be checked. As proposed, we excluded values that are smaller than 0.001 from checking. For the remaining weights, we excluded all bits except for 2 bits in fraction field from checking. As attackers will have access to the pre-trained model only, we simulated their attempts by excluding weights based on the values in the pre-trained model. We verified the pruning accuracy by checking the fine-tuned weight values. If the value change is larger than 0.002 or the sign bit is changed, we considered that the pruning leads to incorrect extraction. Figure~\ref{fig.pruning_efficiency} shows the reduced checking breakdown. The \textit{Weights} bar chart shows the total number of weights correctly pruned. The \textit{Bits} bar chart shows the total number of bits (when considering 32-bit single precision weights) correctly excluded from checking. From the weight checking pruning, we were able to exclude 90\% weights and 85\% bits of weights from checking.

%\subsection{Task Sensitivity}

%BERT models are fine-tuned for various tasks. We evaluated if there is any impact of different tasks for the weight value similarity. We downloaded a BERT-base pre-trained model and fine-tuned it for nine tasks by using Huggingface GLUE benchmark. Then, compared the weight value distances among the nine models. As can be seen in Figure~\ref{fig.weight_diff}, all layers except for the task-dependent last layer show almost zero distances (< 0.002 on average). The last layer showed almost 10$X$ higher differences than the other layers but the weight difference is still around the typical weight value distance between a pre-trained and its fine-tuned. 

\subsection{Applicability to non-Transformer Models}

\begin{figure}[t]
\centering
  \includegraphics[width=0.5\textwidth]{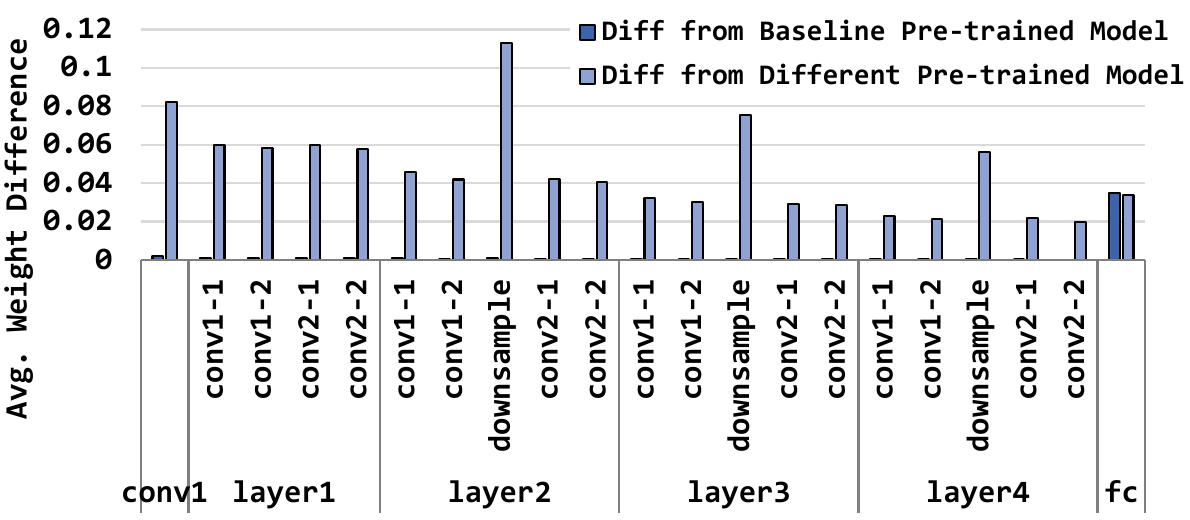}
  \caption{Weight Similarity in a CNN Model (ResNet-18)}
\label{fig.resnet}\vspace{-10pt}
\end{figure}
We believe the weight value similarity is not a unique characteristics of Transformer models. All models that use transfer learning may have the same issue. To verify our intuition, we measured per-layer weight difference of a popular CNN model, ResNet-18, between a fine-tuned model and its baseline pre-trained model as plotted in Figure~\ref{fig.resnet}. We downloaded a pre-trained model from the PyTorch Vision repository~\cite{torch_hub} and fine-tuned it with Hymenoptera database. For a comparison, we trained another ResNet-18 model with the same Hymenoptera database from scratch (without using transfer learning) and compared the weight values with the aforementioned fine-tuned model. The darker blue bar chart in the figure is the weight difference between the fine-tuned model and its baseline pre-trained model. The lighter bar chart is the compared result with a model that is trained from scratch. As clearly noticed, the fine-tuned model has almost zero (< 0.002) weight difference from the baseline pre-trained model in almost all layer, while it shows at least 10$\times$ higher difference from a different model even when they are trained with the same database. We believe this weight similarity is sourced from various parameters used in the initial training such as random seed values. Therefore, we believe this phenomenon will be observed from any models that use transfer learning and hence our weight value similarity-based model extraction attack is applicable for wider range of models.
  
%%%%%%%%%%%%%%%%%%%%%%%%%%%%%% backup data %%%%%%%%%%%%%%%%%%%%%%%%%%%%%%%%%%%%%%%%5
\begin{comment}

\begin{enumerate}
    \item Pytorch-NVIDIA-Bert Base 
            Pretrained/Finetuned
    \item Pytorch-NVIDIA-Bert Large
            Pretrained/Finetuned
    \item Pytorch-Huggingface-Bert Base
            Pretrained/Finetuned
    \item Pytorch-Huggingface-Bert Large
            Pretrained/Finetuned
    \item Pytorch-RoBerta-Large
            Pretrained/Finetuned
    \item Tensorflow-NVIDIA-Bert Large (Tensorflow 2)      Pretrained/Finetuned
    \item Tensorflow-NVIDIA-Bert Base
            Pretrained/Finetuned
    \item Tensorflow-Google-Bert Large (Tensorflow 1.15)   Pretrained
    \item Tensorflow-Google-Bert Base
            Pretrained
    \item Tensorflow-Google-Albert Base/Large (not yet)    Pretrained
\end{enumerate}
\end{comment}

%\section{Discussion}
%\subsection{Countermeasure}
%\hyeran{any thoughts about countermeasure. If space and time permits, we will also do a quick experiment with some countermeasure ideas.}
%\subsection{Approach Generality}

\section{Potential Counter Measures}
For model extraction attacks, we believe that the root cause is consistent execution patterns. %Execution patterns are more critical source than the side-channel interfaces. Thus, some countermeasures like limiting the accesses to performance counters cannot be an ultimate solution. It only makes many other benign researcher's life more difficult because performance counters are important tools for optimizations. As we demonstrated, for simple models like BERT, kernel execution times and sequences are sufficient enough to reveal most of the secrets without detailed architectural hints.     
To break the execution consistency, we showed that some vendor/framework-specific optimizations make notable noises. However, if the noise also has some patterns, as we found from TensorFlow XLA examples, it is only a matter of time that attackers understand the pattern. Therefore, the same optimizations should not be used all the time. Employing bogus kernels or thread blocks have been studied to obfuscate GPU kernel executions~\cite{adwait_hpca20}. However, bogus kernels incur significant performance overhead. To catch both performance and security with one solution, we can consider randomizing the selections of GPU kernels/libraries and usages of various optimizations. Though there are only handful of GPU deep learning libraries (e.g., cuDNN, cuBLAS), an algorithm (e.g., tensor calculation) can be executed with various combinations of library functions. If the combination selection is randomly determined at run time, it would be challenging to extract computation patterns. For example, if one long kernel execution is partitioned into many short kernels, the execution pattern will look very different. Regarding weight value similarity between pre-trained and fine-tuned models, one easiest way is to apply different quantization for fine-tuned models. %For example, pre-trained model may be released with float32 data, while the model is fine-tuned with int8. The actual values may look more different than using the same data type. 
However, due to limited quantization options as discussed in Section~\ref{sec:quant}, it would be eventually understood by the attackers. Instead, we can consider encoding the fine-tuned weights with homomorphic encryption~\cite{homomorphic_1, homomorphic_2} that can be directly computed without decryption. However, the area and performance of homomorphic encryption would be another hurdle to resolve. 

\section{Related Work}
%\hyeran{mostly describe model extraction attacks}

% DeepSniffer~\cite{deepsniffer_asplos20} presented a model extraction attack without any prior information about the victim model. They exploited the static library call sequence, cache miss patterns of individual kernels, and data dependencies among kernels to identify the layer types and the topology of the network. To improve the prediction accuracy, they employed DNNs for model extraction. For the topology extraction, they used LSTM that is used for speech recognition to leverage the classification capability based on time-series inputs. They used GPU performance counters to mimic bus snooping attacks. 

% DeepSteal~\cite{deepsteal}

%\hyeran{add more related work at least 10 papers}

% https://arxiv.org/pdf/1910.12366.pdf
% --> model prediction cloning through exhaustive prediction checking with victim model; attacker knows that the victim uses either bert base or bert large. can't identify the model architecture. definitely they don't consider any noise from vendor and framework usages.

% https://arxiv.org/pdf/2202.06862.pdf
% --> this is not a model extraction model as is but a survey paper that explains all kinds of attack models for pre-training based bert model. good to discuss a little bit. none of the explained attacks leverages weight value similarity between pretrained and fined models.

\subsection{Model Extraction Attack}
Several studies demonstrated various model extraction attacks~\cite{tramer2016stealing, deepsniffer_asplos20, HermesAttack, cacheDNN, weighTheftNoiseInput, deepsteal, yue2021black}. The target models range from basic machine-learning (ML) models like linear regression and support vector machine to most recent neural network models. % and the method varies. 
Tram{\`e}r et al.~\cite{tramer2016stealing} demonstrated an ML model extraction through iterative request transfer to Machine-Learning-as-a-Service servers. Hu et al. and Naghibijouybari et al.~\cite{deepsniffer_asplos20, hoda_ccs18} proposed a framework to extract a black-box CNN model architecture by modeling the mapping between GPU architectural hints and model architecture as a ML recognition problem. Zhu et al.~\cite{HermesAttack} proposed a technique to clone CNN model weights by monitoring and analyzing unencrypted PCIe bus traffic. Yan et al.~\cite{mengjia_security20} introduced a side-channel attack for model extraction based on cache behaviour. Roberts et al.~\cite{weighTheftNoiseInput} demonstrated that attackers can extract victim model weights by injecting random noise inputs to softmax layers. They assume that the attacker is aware of the victim model architecture and has access to softmax layer. Rakin et al.~\cite{deepsteal} leveraged RowHammer~\cite{rowhammer} attack to recover weight values of victim models. For 8-bit weights, they were able to recover almost perfect value with over 4000 rounds of hammerings. %Yue et al.~\cite{yue2021black} investigated the possibility of model extraction attack in sequential recommendation systems, where they first make model extraction attack on black-box recommender models and then perform adversarial attack by using samples generated from the white-box recommender. 
\textbf{These earlier studies mostly targeted CNN models, while we propose a new model extraction attack for Transformer models by leveraging security vulnerabilities of transfer learning.} %However, none of them reveal the severe issues behind the popular pretrain-finetune paradigm. 

\subsection{Attack on Transformer Models}
As Transformer models become more popular due to their state-of-the-art performance in various ML domains, there have been several studies that tackled security concerns about Transformer models. Some studies \cite{krishna2019thieves, DBLP:journals/corr/abs-2108-13873, DBLP:journals/corr/abs-2105-10909} created clone models of a victim BERT model through heuristic testings. By feeding a few random inputs and collecting the predictions of a victim BERT model through online BERT API query services, they were able to train a model that predicts almost similarly (or even better) with the victim model. As their attack goal was not revealing the model architecture, they used a randomly selected BERT pre-trained model. Their experimental results show that the clone model that is built with the same model architecture with the victim model produces more similar predictions. %and fine-tuned the model with the victim model's input-prediction pairs to mimic the victim model's prediction. %Lyu et al.~\cite{} proposed a two-phase model extraction attack targeting at the online service. The proposed attack extracts the model with a few sample queries and labels them using the assumed online serving API. Then they train an extracted model on the resulting data. Secondly, an extra attribute inference model is trained to infer the demographic attributes on any input. 
%Other than security perspectives, several works
Some other studies revealed vulnerabilities of algorithms used by Transformer models~\cite{mahmood2021robustness, wei2021towards, guo2021gradient, joshi2021adversarial}. %examined robustness and vulnerability of Transformer models on adversarial attacks on different granularity. 
They leveraged token-level or sentence-level adversarial samples to attack the victim model in either training phase or inference phase. \textbf{These prior works either focus on the theoretical/empirical analysis on ML characteristics of the Transformer models or reproduce a functionally similar model as the victim online serving model. We provide a comprehensive analysis on the security problems issued by the publicly accessible pre-trained models.} 

\section{Conclusion}
This paper raises a new security concern for transfer-learning models. We show that vendor and framework signature can be leveraged to locate the pre-trained model that the victim model is fine-tuned with. The identified pre-trained model becomes a good source of revealing almost similar weight values in the entire model level. We design a CNN model for the automated model extraction and propose methods to recover actual weights.

%\section*{ACKNOWLEDGMENTS}
%This document is derived from previous conferences, in particular MICRO 2013, ASPLOS 2015, MICRO 2015, MICRO 2016, MICRO 2017, MICRO 2018, MICRO 2019, MICRO 2020 and MICRO 2021, as well as SIGARCH/TCCA's Recommended Best Practices for the Conference Reviewing Process.

%%%%%%% -- PAPER CONTENT ENDS -- %%%%%%%%

%%%%%%%%% -- BIB STYLE AND FILE -- %%%%%%%%
\bibliographystyle{IEEEtran}
\bibliography{bibliograph}
%%%%%%%%%%%%%%%%%%%%%%%%%%%%%%%%%%%%

% that's all folks
\end{document}